\documentclass[fleqn,usenatbib]{mnras}
\usepackage[T1]{fontenc}
\usepackage{ae,aecompl}
\usepackage{graphicx}	
\usepackage{amsmath}	
\usepackage{amssymb}	

\usepackage{textcomp}
\usepackage{multicol}  
\usepackage{bm}		   
\usepackage{pdflscape} 
\usepackage{color}     
\title[Testing isotropy of CMB polarization maps]{Testing Statistical Isotropy in Cosmic Microwave Background Polarization maps}

\author[P. K. Rath et al.]{Pranati K. Rath,$^{1,2}$\thanks{Email: pranati@ihep.ac.cn}
Pramoda Kr. Samal,$^{3}$\thanks{Email: pksamal@iopb.res.in}
Srikanta Panda$^{3}$, Debesh D. Mishra$^{4}$,
\newauthor{Pavan K. Aluri$^{5,6}$}
\\
$^{1}$Institute of High Energy Physics, Chinese Academy of Sciences, Beijing 100049, China\\
$^{2}$Theoretical Physics Center for Science Facilities, Chinese Academy of Sciences, Beijing 100049, China\\
$^{3}$Department of Physics, Utkal University, Bhubaneswar, Odisha, 751004, India\\
$^{4}$State Key Laboratory of Advanced Technology for Materials Synthesis and Processing, Wuhan University of Technology, Wuhan 430070, China\\
$^{5}$Korea Institute for Advanced Study, Quantum Universe Center,
      85 Hoegi-ro, Dongdaemun-gu, Seoul - 02455, Republic of Korea\\
$^{6}$Korea Astronomy and Space Science Institute,
      776 Daedeokdae-ro, Yuseong-gu, Daejeon - 305348, Republic of Korea
}

\date{Accepted XXX. Received YYY; in original form ZZZ}

\pubyear{2017}

\begin{document}
\label{firstpage}
\pagerange{\pageref{firstpage}--\pageref{lastpage}}
\maketitle

\begin{abstract}
We apply our symmetry based \emph{Power tensor} technique to test conformity of PLANCK Polarization
maps with statistical isotropy.
On a wide range of angular scales ($l=40-150$), our preliminary analysis detects many statistically
anisotropic multipoles in foreground cleaned full sky PLANCK polarization maps viz.,
COMMANDER and NILC. We also study the effect of residual foregrounds that may still be present
in the galactic plane using both common UPB77 polarization mask, as well as the individual component
separation method specific polarization masks. However some of the statistically anisotropic modes
still persist, albeit significantly in NILC map. We further probed the data for any coherent
alignments across multipoles in several bins from the chosen multipole range.
\end{abstract}


\section{Introduction}

The standard cosmological model is based on the postulate that the universe is homogeneous and
isotropic on large distance scales. However, there exist many  observations which suggest that
this postulate is violated. The first indication of statistical anisotropy came from the analysis
of radio polarization from distant radio galaxies \citep{Birch1982, Jain1999, Jain2004}. Those
authors found that the polarization offsets, after eliminating the effect of Faraday rotation
show a dipole pattern on the sky and the dipole axis points towards the Virgo cluster of galaxies,
very close to the direction of the CMB dipole \citep{Jain1999}. The optical polarizations of
distant quasars shows alignment over very large distance scales \citep{Hutsemekers1998, Hutsemekers2001}.
The distance scale of alignment is found to be of the order of 1 Gpc \citep{Jain2004}, and the
axis again points close to the CMB dipole axis.
The CMB quadrupole and octopole also indicates a preferred direction pointing towards the Virgo
cluster \citep{Copi2004, Ralston2004, Costa2004, Schwarz2004, Bielewicz2005, Samal2008}.
This phenomenon where several axes from various data sets broadly indicate the same direction
has been called \emph{Virgo alignment puzzle} \citep{Jain2004}.
Besides the Virgo alignment, there also exists statistically significant signals of anisotropy
in CMB temperature data viz., hemispherical power asymmetry \citep{Eriksen2004a, Prunet2005,
Hansen2009, Hanson2009, Hoftuft2009, Rath2013, Akrami2014, Aiola2015, Pranati2015},
parity asymmetry \citep{Land2005, Kim2010, Kim2011, Aluri2012, Zhao2014, Aluri2017} and a region
of significant temperature decrement known as cold spot \citep{Vielva2004, Cruz2005, Cruz2006,
Cruz2008, Zhao2013, Nadathur2014, Aluri2016}.

Large angle CMB anisotropies have been a subject of several studies in the cosmology literature
- see for example \citet{wmap7yranom,Planck2013iso,Planck2015iso} for an evaluation of some
of the prominent large angle anomalies by WMAP and PLANCK collaborations.
General methods have been developed to test any violation of statistical isotropy in the CMB data
\citep{Hajian2003, Hajian2006, Copi2004, Copi2006, Samal2008, Samal2009}.
In \citet{Samal2008, Samal2009}, a symmetry based method for testing the isotropy of CMB
temperature data called \emph{Power tensor} was proposed.
The method is based on identifying invariants corresponding to each multipole using the
Power tensor matrix defined as
\begin{equation}
A_{ij} (l) = \frac{1}{l(l+1)(2l+1)}\sum_{m,m'} a^*_{lm}(J_{i} J_{j})_{mm'} a_{lm'}\,.
\label{eq:pt}
\end{equation} 
Here $J_i (i = 1, 2, 3)$ are the angular momentum operators in spin$-l$ representation.
The sum of the three eigenvalues reproduce the usual angular power spectrum, $C_l$.
Statistical nature of CMB anisotropies lead to fluctuations in the eigenvalues of Power tensor
about their expected value of $C_l/3$ in a given realization.
The significance of any deviation from isotropy is measured using an invariant combination
of normalized eigenvalues of the Power tensor called \emph{Power entropy}
\citep{Samal2008,Samal2009,Rath2015}.

In this paper, we probe statistical anisotropy, if any, in the PLANCK CMB polarization
data using our symmetry based method, and its possible relation to the existing anomalies
of CMB temperature sky. A violation of statistical isotropy can arise due to a variety
of sources, for example, from residual contamination due to foregrounds, beam systematics,
inhomogeneous noise, etc. Here we do not make any attempt to find the exact cause of
breakdown of isotropy, if found eventually in our analysis.

The paper is organized as follows. In section~\ref{sec:revpol}, we briefly review CMB
polarization convention. Then the statistics used to analyze the data are defined in
section~\ref{sec:stat}. The data used in the present work is described in section~\ref{sec:data},
and our results are discussed in section~\ref{sec:results}. There, we first present
the results from analyzing full sky PLANCK CMB polarization maps. Then we go on to discuss
the effect of foreground residuals if any on our results by employing different galactic
masks, also provided by PLANCK team. We also present results from our tests for any
coherent alignments among various multipoles. Finally, the work is summarized in
section~\ref{sec:con}.

\section{CMB polarization maps}
\label{sec:revpol}
The full sky CMB signal is described by Stokes parameters $I$, $Q$, $U$ and $V$. The Stokes
parameter $I$ represents the temperature field, and the Stokes parameters $(Q,U)$ represent
the linear polarization field. In the standard model of cosmology, the CMB temperature
fluctuations are expected to be an isotropic Gaussian random field, and are conventionally
expanded in terms of the spherical harmonics as 
\begin{equation}
T(\hat n) = \sum_{lm} a_{lm}Y_{lm}(\hat n)\,,
\end{equation}
where $a_{lm}$ are the coefficients of expansion. The CMB polarization is induced by Thomson
scattering of CMB photons at the last scattering surface. The WMAP and PLANCK teams provided
maps of CMB polarization in terms of the Stokes parameters $Q$ and $U$, though with low signal
to noise ratio (SNR). The Stokes parameter $V$, which describes circular polarization is
ignored as it cannot be generated through Thomson scattering.

From now onwards we use the notation of \citet{Zaldarriaga1997} for the polarization fields.
Instead of the Stokes parameters $Q$ and $U$, it is useful to employ special combinations of
these parameters as $X_\pm=Q \pm i U$, which transform as spin-$2$ fields under coordinate
rotation. Under a rotation by an angle $\phi$ of the co-ordinate frame in which the polarization
vector is defined, the combinations $X_\pm$ behave as spin $\pm 2$ fields viz.,
\begin{equation}
(Q\pm i U)'(\hat{n}) = e^{\mp 2 i \phi}( Q\pm i U)(\hat{n})\,.
\end{equation} 
Analogous to the expansion of temperature field in terms of the spherical harmonics,
$Y_{lm}(\hat n)$, there also exists a set of spin-$s$ spherical harmonics $_{s}Y_{lm}(\hat n)$,
in terms of which one can expand a spin-$s$ function on a sphere. We can, therefore,
expand $X_\pm(\hat n)=( Q\pm i U)(\hat{n})$ in terms of the spin-$2$ spherical harmonics as
\begin{equation}
X_\pm(\hat{n})=\sum_{lm} a_{\pm 2,lm} \ _{\pm 2}Y_{lm}(\hat{n})\,.
\label{spin2}
\end{equation}
Owing to the real nature of $Q$ and $U$ parameters, the expansion coefficients satisfy
the condition : $a^*_{-2,lm} = a_{2,l-m}$. Using the following identities
\begin{eqnarray}
\eth \ _{s}Y_{lm} &=& \sqrt{(l-s)(l+s+1)} \ _{s+1}Y_{lm}\,,\\
\bar{\eth}\ _{s}Y_{lm} &=& -\sqrt{(l+s)(l-s+1)} \ _{s-1}Y_{lm}\,.
\end{eqnarray}
for the spin raising and lowering operators : $\eth$ and $\bar{\eth}$, spin-$0$ objects
can be constructed from spin-$2$ fields as
\begin{eqnarray}
\bar{\eth}^2(Q+ i U)(\hat{n})&=&\sum_{lm}\sqrt{\frac{(l+2)!}{(l-2)!}} \ a_{2,lm}\ Y_{lm}(\hat{n})\,,\\
\eth^2(Q- i U)(\hat{n})&=&\sum_{lm}\sqrt{\frac{(l+2)!}{(l-2)!}} \ a_{-2,lm}\ Y_{lm}(\hat{n})\,.
\end{eqnarray}

Finally, the rotationally invariant polarization fields are defined as
\begin{eqnarray}
\tilde{E}(\hat{n})& = &-\frac{1}{2}\left[\bar{\eth}^2(Q+ i U)+\eth^2(Q- i U)\right]\nonumber\\
& = &\sum_{lm} \tilde{a}^E_{lm}Y_{lm}(\hat{n})\,,
\end{eqnarray}
\begin{eqnarray}
\tilde{B}(\hat{n}) & = &-\frac{1}{2i}\left[\bar{\eth}^2(Q+ i U)-\eth^2(Q- i U)\right]\nonumber\\
& = &\sum_{lm} \tilde{a}^B_{lm}Y_{lm}(\hat{n})\,
\end{eqnarray}
where $\tilde{a}^{E/B}_{lm}=\sqrt{(l+2)!/(l-2)!} \ a^{E/B}_{lm}$, and $a^E_{lm}$ and
$a^B_{lm}$ are the spherical harmonic coefficients of ``$E$-mode'' and ``$B$-mode''
polarization fields. Note that often $E/B$ harmonic coefficients (and consequently
the $E/B$ fields) are defined without the extra $\sqrt{(l+2)!/(l-2)!}$ factor.
These $E/B$ spherical harmonic coefficients are given by linear combinations of
spin-$2$ spherical harmonic coefficients of Stokes $Q/U$ polarization fields as
\begin{equation}
 a^E_{lm}=-\frac{1}{2}\left( a_{2,lm}+ a_{-2,lm}\right)\,,
\label{eq:Epol}
\end{equation}
\begin{equation}
 a^B_{lm}=-\frac{1}{2i}\left( a_{2,lm}- a_{-2,lm}\right)\,.
\label{eq:Bpol}
\end{equation}

Here we restrict our attention to the $E(\hat n)$ field (ie., $a^E_{lm}$ coefficients),
to study statistical isotropy of PLANCK polarization maps.

\section{Statistics}
\label{sec:stat}
The angular orientation of each mode is given by a unique orthonormal frame $e_k^{\alpha}(l)$
and rotationally invariant singular values $\Lambda_{\alpha}(l)$ of the Power tensor, $A(l)$, 
defined in Eq.~(\ref{eq:pt}). Here $k=1,2,3$ denote the components of the frame vector
${\bf e}^\alpha(l)$ and $\alpha=1,2,3$ stands for the singular value index.
In terms of these quantities, the Power tensor matrix for each multipole `$l$' can be
expressed as
\begin{equation}
A_{ij}(l) = \sum_{\alpha} e_{i}^{\alpha}(\Lambda^{\alpha})^2 e_j^{\alpha*}\,.
\end{equation}
We do not explicitly display the index $l$ when it is obvious.
We refer to the eigenvector corresponding to the largest eigenvalue of the Power tensor
as \emph{principal eigenvector} (PEV), and is taken to be the anisotropy axis of that
multipole. The preferred direction represented by a PEV of any multipole is quantified by
parametrizing the dispersion of eigenvalues using \emph{Power entropy} that is defined as
\begin{equation}
S_{p} (l) = -\sum_{\alpha} \lambda^{\alpha} \log(\lambda^{\alpha}) \,
\label{eq:powent} 
\end{equation}   
where $\lambda^{\alpha} = (\Lambda^{\alpha})^2/\sum_{\beta}(\Lambda^{\beta})^2$. In the ideal
case of isotropy, where all the three eigenvalues are degenerate and equal to ${C_l}/3$, we
have maximum Power entropy, $S_p \rightarrow \log(3)$. In the case of a \emph{pure state},
where one of the eigenvalues contains the total power and other two vanishes, it leads to
vanishing Power entropy, $S_p \rightarrow 0$. So for our observational data, the range of
Power entropy is $0\le S_p \le \log(3)$. Hence low Power entropy in data, compared to consistently
generated concordance model simulations, is a measure of isotropy violation in the data.

Now, a common alignment vector using PEVs for a set of multipoles or range of multipoles
can be calculated using what is called an \emph{Alignment entropy}, defined by
\begin{equation}
S_{X} = -Tr(\rho_X \log(\rho_X))\,,
\label{eq:alignent}
\end{equation}
where $\rho_X = {X}/{Tr(X)}$ is the normalized $3\times 3$ matrix `$X$' that is referred
to as \emph{Alignment tensor}. It is given by
\begin{equation}
X_{ij} = \sum_{l_{min}}^{l_{max}} e^i(l)e^j(l) \,,
\label{eq:alignten}
\end{equation}
where ${\bf e}(l)$ is the PEV of a multipole, $l$. $Tr(X)$ denotes trace of $X$.
An unusually low value of $S_{X}$ compared to $\log(3)$ confirms violation of isotropy
over a wider multipole range.
We note that the Power entropy and the Alignment entropy are independent of each other.

The significance of statistical anisotropy is determined by comparing the data statistic
value with that of simulations and the significance is quoted by the $P$-value.
A $P$-value is defined as the probability that a random realization may yield a statistic
smaller than that seen in data.
The effective probability for a collection of PEVs with respective $P$-values less than
a reference probability `$\mathbb{P}$', is estimated using the binomial distribution of
\emph{pass} and \emph{fail} outcomes. The probability to encounter $k$ instances of
passing defined by probability $\mathbb{P}$ in $n$ trials is
\begin{equation}
f(k|n,\mathbb{P})= \frac{\mathbb{P}^{k}(1-\mathbb{P})^{n-k}n \,!}{(n-k)\,! k\,!}
\label{eq:prob}
\end{equation}
In assessing many $P$-values, we report the cumulative binomial probabilities as
\begin{equation}
f(k \ge k_{*}|n,\mathbb{P})= \sum_{k=k_{*}}^{n} f(k|n,\mathbb{P})
\label{eq:cumprob}
\end{equation}
which is the probability to see $k_*$ or more instances of passing among $n$ trials
defined by a threshold probability $\mathbb{P}$.

\section{Data Used}
\label{sec:data}
The PLANCK team has provided four foreground reduced CMB polarization maps referred to as
COMMNDER, NILC, SMICA and SEVEM \citep{Planck2015cmb} maps, named after the component
separation procedure used. Out of these four, we study only the COMMNDER and NILC as they
are full sky maps which are suitable for our study. 
The other clean CMB maps viz., SMICA and SEVEM polarization solutions have a portion of
the sky removed, particularly in the galactic plane. Hence we will not use them in our
analysis. All these cleaned polarization CMB maps were estimated using all the frequency
channels aboard PLANCK that are sensitive to polarization (from $30$ to $353$ GHz).
Due to the presence of significant noise in the polarization maps and also due to the
possible residual foregrounds that may still be present even after cleaning, any signature
of large scale isotropy breakdown has to be interpreted with care. Here we use half-ring
half-difference (HRHD) maps \citep{Planck2015cmb} as noise proxy in our analysis. 
The publicly available polarization maps don't include low-$l$ up to $l = 40$ owing to
systematics \citep{Planck2015cmb}. 
Consequently, we only analyze multipoles $l \geq 40$. As noise contribute dominantly to
the polarization maps from PLANCK at high-$l$, we restrict our analysis up to $l=150$.

\begin{figure*}
\centering
\includegraphics[width=0.32\textwidth]{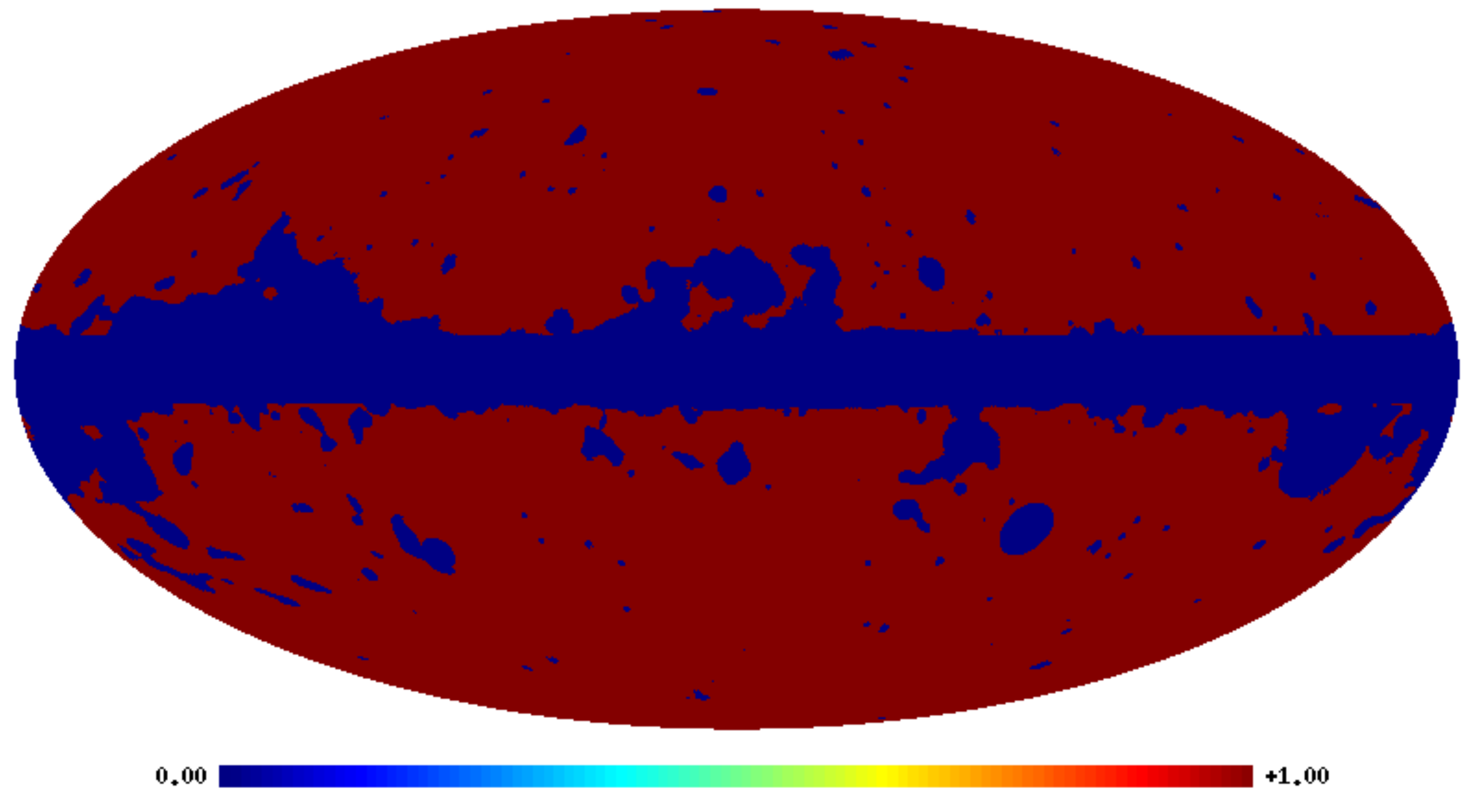}
~
\includegraphics[width=0.32\textwidth]{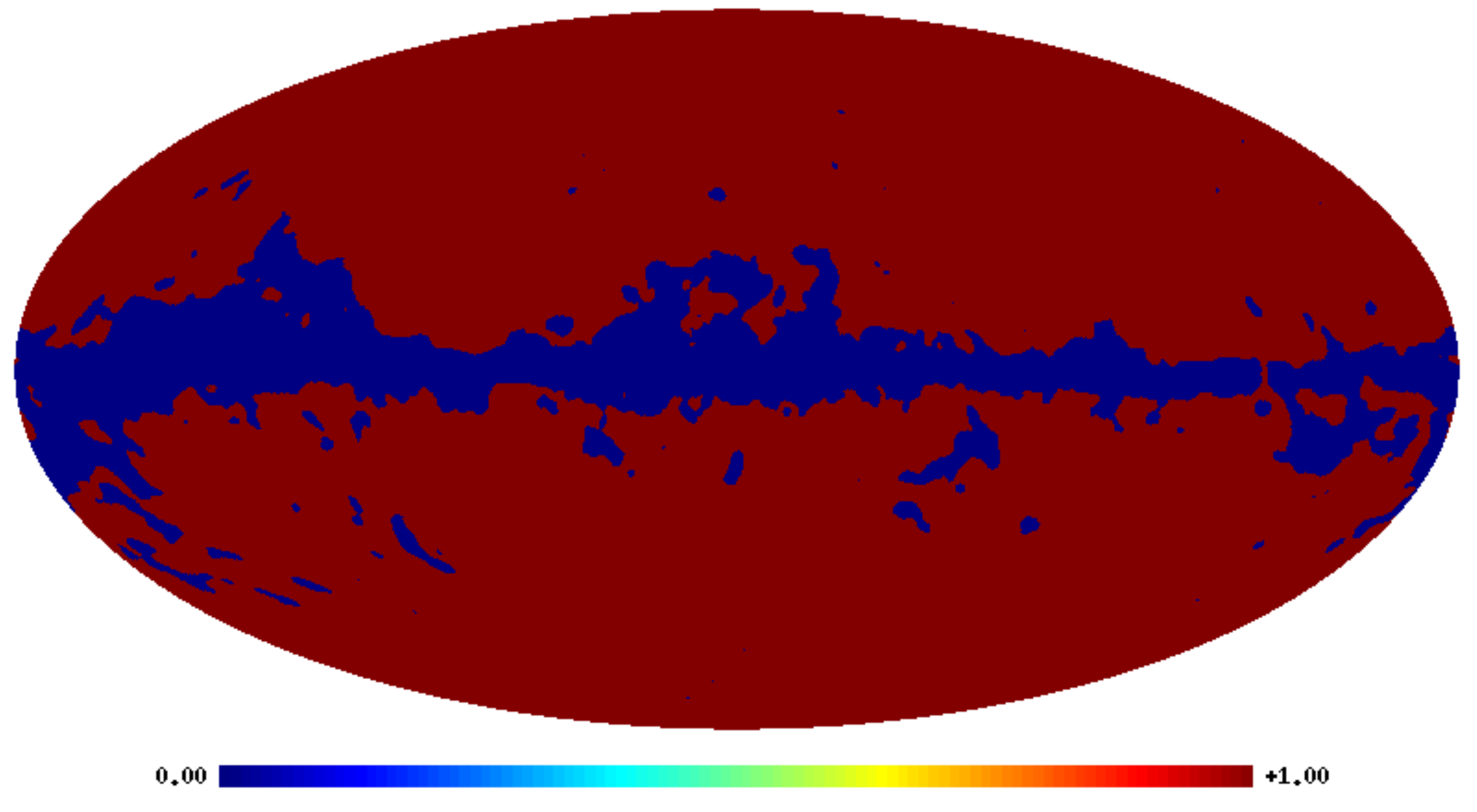}
~
\includegraphics[width=0.32\textwidth]{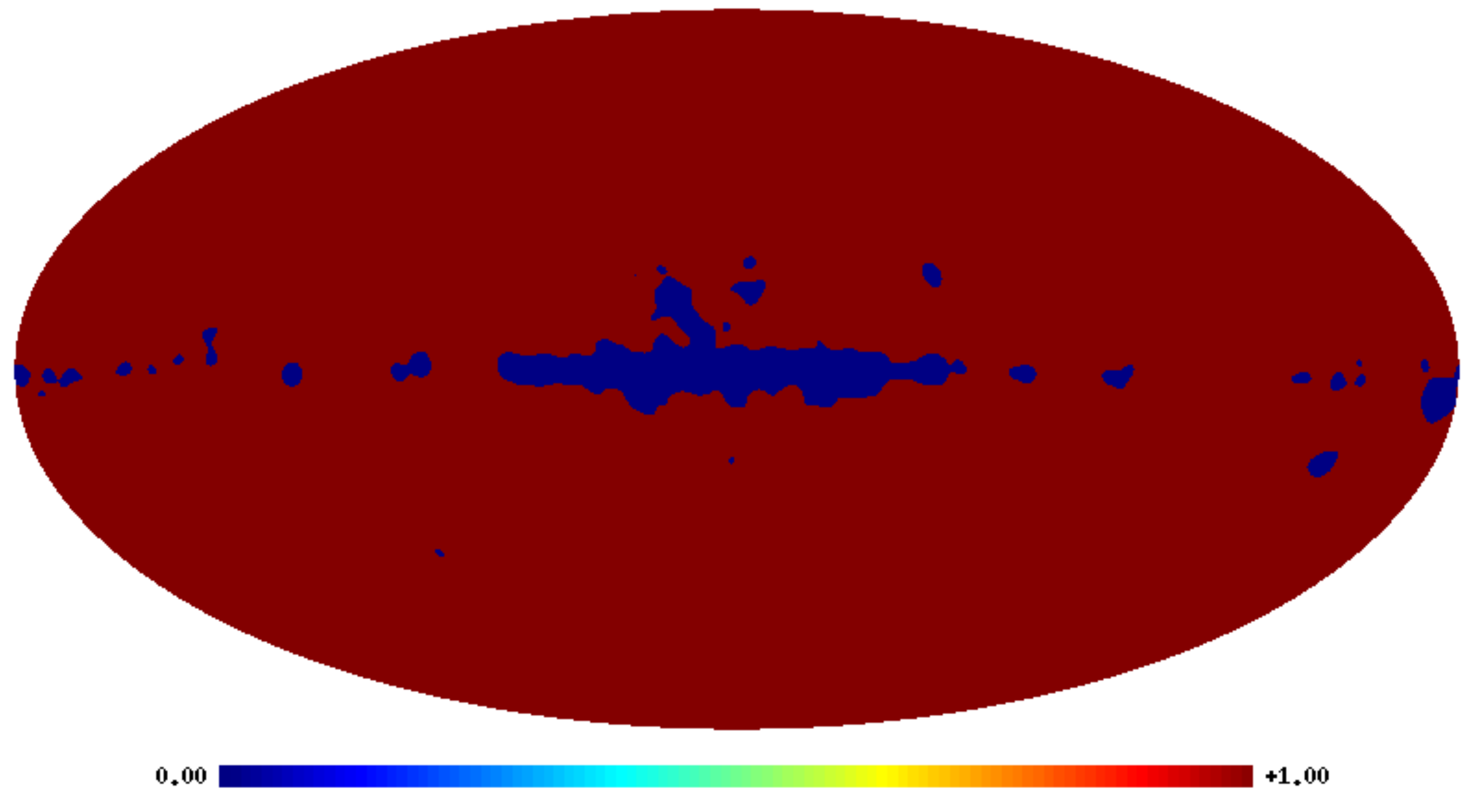}
\caption{\emph{Left} : $UPB77$ common polarization analysis mask, \emph{Middle} : COMMANDER polarization map
         ($PMCMDR$), \emph{Right} : NILC polarization mask ($PMNILC$). All these masks
         are available at $N_{side} = 1024$.}
\label{fig:mask-pol}
\end{figure*}

To start with, we extract the $E$-mode polarization map from the full sky Stokes $Q$ and $U$
maps from PLANCK available at a HEALPix \citep{Gorski2005} resolution of $N_{side}=1024$
and have a beam resolution given by a Gaussian beam of $FWHM=10'$ (arcmin). 
We first analyze the full sky COMMNDER and NILC foreground cleaned PLANCK $E$-mode
polarization maps thus obtained.
Subsequently we generate foreground-residual minimized full sky CMB $E$-mode data maps by
following the procedure described in \citet{Samal2010,Pranati2013,Pranati2015}.
For the purposes of masking, we use the common polarization mask $UPB77$, as well as the
polarization masks specific to each component separation method employed by PLANCK \citep{Planck2015cmb}.
These masks are shown in Fig.~(\ref{fig:mask-pol}). The common polarization mask, $UPB77$
cover $\approx 77\%$ of the sky whereas the COMMANDER and NILC polarization masks cover
$\approx83\%$ and $96\%$ of the sky respectively. From now on we abbreviate the COMMANDER
polarization mask as $PMCMDR$ and the NILC polarization mask as $PMNILC$.

A full sky CMB $E$-mode data map with minimized residual foregrounds is generated following
the steps listed below :
\begin{enumerate}
\item We first generate a full sky CMB $IQU$ random realization using the best fit theoretical
angular power spectrum ($C^{th}_l$). The best fit $C^{th}_l$ are generated using PLANCK 2015
cosmological parameters \citep{Planckpara2016a,Planckpara2016b} as input to CAMB
software \citep{Lewis2000, Howlett2012}. The values of cosmological parameters from PLANCK 2015
results that we used are baryon matter density $\Omega_{b}h^2 = 0.0222$, cold dark matter
density $\Omega_{c}h^2 = 0.1203$, neutrino energy density $\Omega_{\nu}h^2 = 0.00064$, 
cosmological constant density fraction $\Omega_{\Lambda}= 0.6823$, Hubble parameter $H_{0}$
with $h=0.6712$, scalar spectral index of the primordial power spectrum $n_{s} = 0.96$,
amplitude of primordial power spectrum $A_{s} = 2.09\times 10^{-9}$, and reionization optical
depth $\tau = 0.078$. The best fit theoretical angular power spectum, $C^{th}_l$, thus obtained
from CAMB using these parameters, is employed to generate random realizations of CMB sky using
HEALPix. A CMB $IQU$ map is generated with a Gaussian beam of $FWHM=10'$ (arcmin)
at $N_{side}=1024$.

\item A cosine filter is then applied on the simulated CMB $IQU$ random realization obtained
in step~(i), to remove the large angular scales following \citet{Planck2015cmb}. The cosine
filter is defined as
\begin{equation}
w_l = 
\begin{cases}
   0, & \text{}\ l<l_1 \\
   \frac{1}{2}\left[1-\cos\left(\pi\frac{l-l_1}{l_2-l_1}\right) \right], & \text{}\ l_1 \le l \le l_2 \\
   1, & \text{}\ l_2 < l,
\end{cases}
\label{eq:cosbeam}
\end{equation}
where $l_1=20$ and $l_2=40$.

\item The $IQU$ noise proxies viz., the HRHD maps corresponding to COMMANDER/NILC component
separation method are added to the filtered CMB $IQU$ realization obtained in step~(ii).
Note that the cosine filter given by Eq.~(\ref{eq:cosbeam}) is already applied to the HRHD
polarization data maps that are made publicly available. This results in a random CMB
realization with noise levels similar to data.

\item Now, the inverse of polarization masks shown in Fig.~(\ref{fig:mask-pol}) are applied
to the random CMB $IQU$ realization added with an HRHD map.
By doing so, the resultant map will have signal only in the galactic region (and few other
regions), with rest of the sky set to zero.
Next, we apply the polarization masks of  Fig.~(\ref{fig:mask-pol}), as they are, on the
CMB data $IQU$ map to remove the potentially contaminated galactic region. 
Finally the two pieces that have complementary regions masked are added together to construct
a full sky CMB $IQU$ data map at $N_{side}=1024$.

\item Subsequently, we derive the $E$-mode polarization map from this composite data $IQU$
map at $N_{side}=1024$.
\end{enumerate}
Thus we have effectively minimized the residual contamination in the data CMB polarization map.

Due to random filling of the masked regions, one will get a (slightly) different value for
the statistic, compared to the \emph{true} sky. Hence our data statistic is taken as average
value of the same quantity derived from $100$ such \emph{filled} full sky COMMANDER and
NILC $IQU$ maps.

The significance of isotropy violation is estimated by comparing the (average) data statistic
value with $4000$ random CMB realizations (added with HRHD noise maps). Since we have different
HRHD maps for COMAMNDER and NILC maps, we generate two sets of $4000$ random realizations.
Simulations of CMB $IQU$ maps with noise levels similar to data are generated following step
(i) - (iii) described above. We then extract the $E$-mode polarization map from these realistic
$IQU$ maps. We note that the temperature realizations generated here are only a by-product and
have no use for us. So any operation like masking or filtering performed on $I$ map, together
with $Q$ and $U$ maps, has no relevance to our study.
 
\section{Results}\label{sec:results}
First we analyze the \emph{full sky} PLANCK $E$-mode polarization maps as obtained from cleaned
CMB data $IQU$ map. Later we use the \emph{filled} data maps to understand the effect of residual
contamination that may still be present in the recovered CMB sky. Recall that in order to minimize
likely foreground bias on our results, we filled the masked regions shown in Fig.~(\ref{fig:mask-pol}),
with a filtered random isotropic realization added with an HRHD map.

\subsection{Power entropy vis-a-vis Axiality of multipoles}
\label{subsec:poweren}
\subsubsection{Full sky analysis}
\begin{figure*}
\centering
\includegraphics[width=0.98\textwidth]{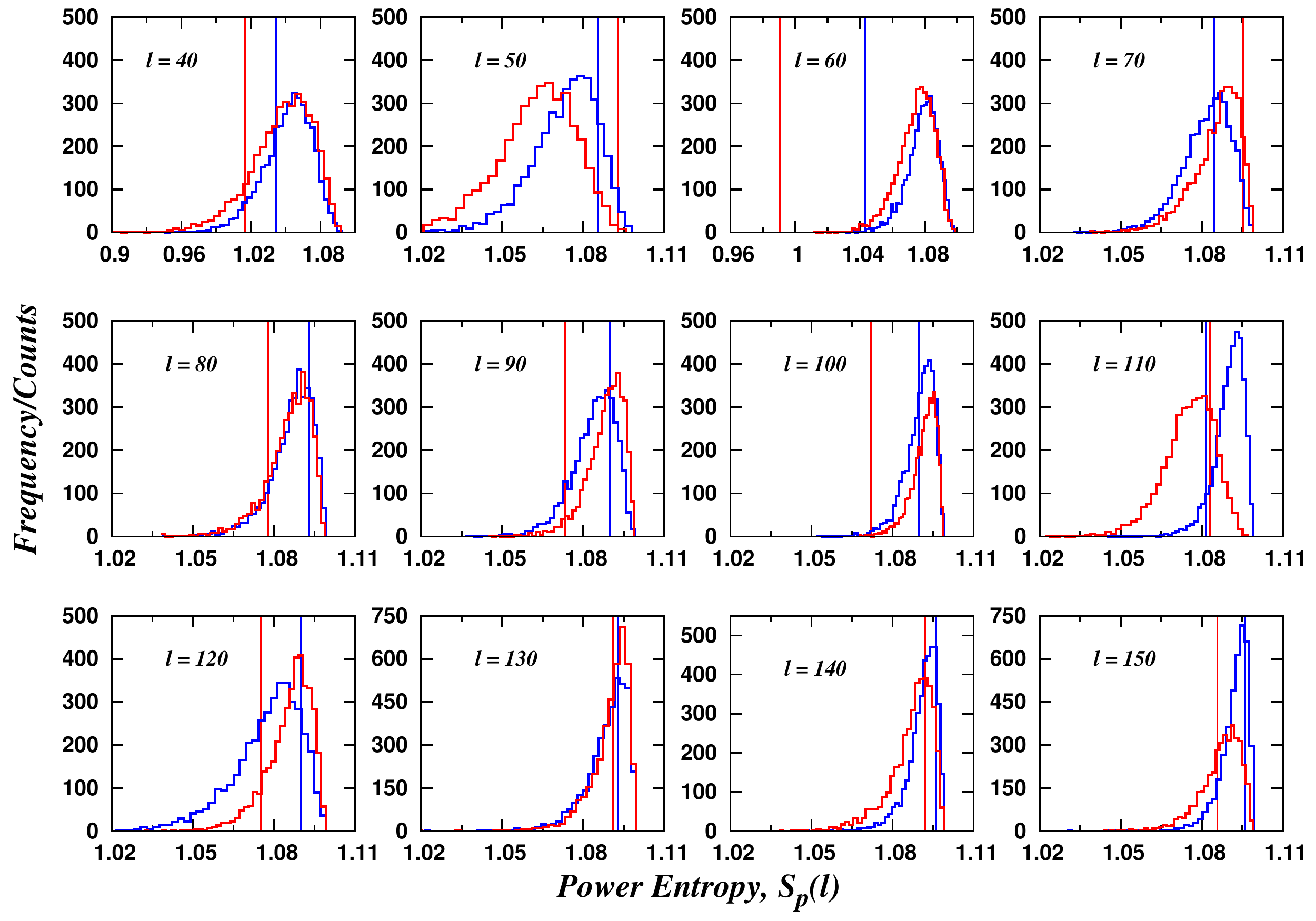}
\caption{Empirical distribution of Power entropy `$S_p$', obtained from $4000$ simulated
         $E$-maps with appropriate noise, for the multipole range $40\le l \le 150$, but
         shown at intervals of $10$ multipoles. The \emph{blue} and \emph{red} color
         histograms in each plot correspond to COMMANDER and NILC simulations. Similarly
         the vertical lines in respective colors indicate the data value from full sky
         $E$-mode polarization maps obtained from COMMANDER and NILC Stokes $Q/U$ maps
         as provided.}
\label{fig:hist}
\end{figure*}

Here we study statistical anisotropy of the multipoles in PLANCK $E$-mode polarization
map as obtained from full sky Stokes $Q/U$ CMB maps, derived using COMMANDER and NILC
cleaning procedures.
The Power entropy, $S_p$, is computed for each multipole in the chosen multipole range
$l=40-150$ from the $E$-mode polarization maps using Eq.~(\ref{eq:powent}).
The statistical significance of the Power entropy values thus computed from data are
studied using $4000$ isotropic random CMB $E$-mode polarization maps that are appropriately
filtered (see Eq.~(\ref{eq:cosbeam}))
and added with the noise proxy of the data i.e., half-ring half-difference (HRHD) maps of
respective component separation methods.
Fig.~(\ref{fig:hist}) shows the null distribution of Power entropy for the multipoles in
the range $l = 40-150$, but at an interval of $10$ multipoles  i.e., for $l=40,50,60\cdots150$
for brevity. The two histograms in each plot corresponding to the two data sets (component
separation maps) used in the analysis and the two vertical lines indicate the respective
data values.

\begin{figure*}
    \centering
\includegraphics[width=0.95\textwidth]{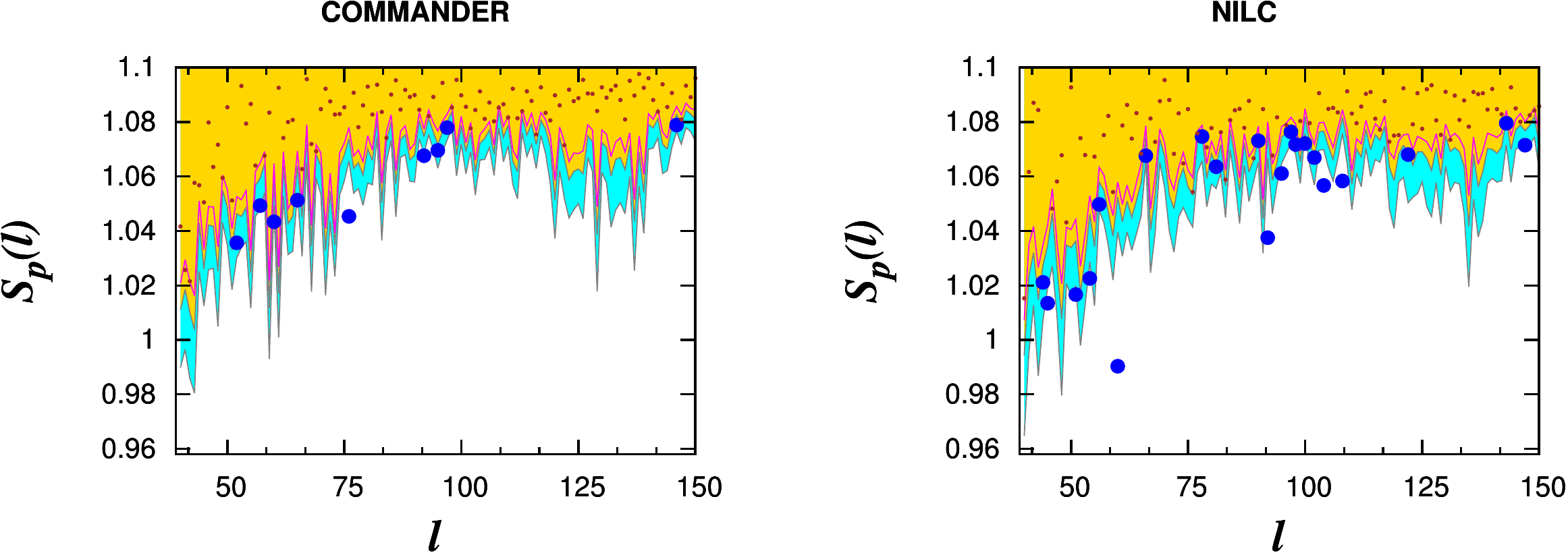}
    \caption{Power entropy ($S_p$) values from full sky COMMANDER (\emph{left}) and
             NILC (\emph{right}) $E$-polarization maps over the range $l=40-150$ are
             shown here.
             The data $S_p$ values in each plot are shown as \emph{red} and \emph{blue}
             dots, where the \emph{blue} points highlight those multioles whose Power
             entropy has a $P$-value less than $5\%$.
             The magenta line, gold and cyan bands represent $90\%$, $95\%$ and $99\%$
             confidence contours estimated from $4000$ simulations.}
    \label{fig:counterplot1}
\end{figure*}

In Fig.~(\ref{fig:counterplot1}), we show the Power entropy, $S_p$, values from data
for all the multipoles in the range $l=40-150$. The data values are denoted by
\emph{red} and \emph{blue} points, where the blue ones correspond to those multipoles
whose Power entropy has a probability of $P\leq 5\%$.
Also plotted are $90\%$, $95\%$ and $99\%$ confidence contours of distribution of Power entropy,
as obtained using 4000 simulations, with a \emph{magenta} line, \emph{gold} and \emph{cyan} colour
bands respectively. The rugged nature of the distribution can be understood from the fact
that we used the same HRHD map in our simulations to mimic the data noise levels.

The list of multipoles which are found to be statistically anisotropic are listed
in second column of Table~(\ref{tab:tab1}) corresponding to each PLANCK component
separated polarization map we used.
We find that there are $k_{data}=9$ and $21$ number of multipoles which have $P \le 0.05$
in the range we analyzed in COMMANDER and NILC $E$-mode polarization maps respectively.
The total number of independent trials for the range $40 \le l \le 150$ is $111$.
Following Eq.~(\ref{eq:cumprob}), we can compute the cumulative probability of finding
$k_{data}$ or more instances of statistically anisotropic modes among all the modes analyzed.
Note that we defined the criteria of pass or fail with a reference probability of $\mathbb{P}=0.05$.
Thus, from the binomial distribution, the {\it cumulative} probability, $f(k \geq k_{data}|111, \mathbb{P}=0.05)$
are found to be $0.104$ and $1.390 \times 10^{-7}$, respectively for COMMANDER
and NILC maps. These are also listed in Table~(\ref{tab:tab1}), in the \emph{third} column.
As is obvious, we find that the cumulative probability for the observed Power entropy in NILC
$E$-mode polarization map is very small compared to that of COMMANDER map. This is so owing
to more number of multipoles being axial in NILC map at the level of $2\sigma$ or more
than in COMMANDER map.

\begin{table}
\scriptsize
\centering
\begin{tabular}{c l c}
\hline
\hline
Map & Multipoles & Cumulative \\
    &            & Probability\\
\hline
\hline
 & \\
COMMANDER   & 52, 57, 60, 65, 76, 92, 95, & $ 0.104 $ \\
              & 97, 146                     & \\
  &  & \\
NILC        & 44, 45, 51, 54, 56, 60, 66,  & $1.390 \times 10^{-7}$  \\
              & 78, 81, 90, 92, 95, 97, 98,  & \\
              & 100, 102, 104, 108, 122, & \\
              & 143, 147                     & \\
  & & \\
\hline
\hline
\end{tabular}
\caption{List of multipoles with probability $P \le 0.05$ for Power entropy from PLANCK's 
	full sky cleaned $E$-mode polarization maps as indicated are given in \emph{second} column.
	The cumulative probability for finding the observed number of statistically anisotropic modes
	in individual cleaned maps with $P\leq0.05$ are furnished in the \emph{third} column.}
\label{tab:tab1}
\end{table}

From Fig.~(\ref{fig:counterplot1}) or Table~(\ref{tab:tab1}) we see that there is a significant
indication of violation of statistical isotropy in the PLANCK polarization maps we studied.
One may argue that this may be arising due to residual foregrounds that are potentially present
in the cleaned maps. In the next section we will try to minimize the effect of this residual
contamination following the procedure described in section~\ref{sec:data} and re-evaluate the significances
reported here.

\subsubsection{Understanding the effect of galactic residuals}
In this section, we present the result obtained from the full sky $E$-mode data polarization
maps constructed as discussed in section~\ref{sec:data}.
We first apply the common $UPB77$ mask on the cleaned PLANCK $IQU$ polarization maps and 
construct the full sky CMB data $IQU$ maps by filling the masked region with a filtered
random CMB realization (Eq.~(\ref{eq:cosbeam}) added with HRHD noise map. We then extract
full sky $E$-mode polarization maps from the filled $Q/U$ maps, and use these to study the
statistical isotropy of multipoles in the range $l=40-150$.
We also use component separation specific polarization masks $PMCMDR$ and $PMNILC$ corresponding
to COMMANDER and NILC procedures respectively to construct full sky $E$-mode polarization maps
following the same procedure.
The exercise is then repeated with these maps to reassess the statistics presented before.
As explained in section~\ref{sec:data}, the data statistic is taken as average value of
the statistic derived from $100$ such randomly filled full sky COMMANDER and NILC $IQU$ maps.
We note that such a filling procedure is expected to reduce any signal of statistical anisotropy
if present, because of the randomizing effect of the filling.

The list of multipoles having $P \le 0.05$ in various cases considered are given in
\emph{second} column of Table~(\ref{tab:tab2}).
Now, using the common UPB77 mask for filling, we find that there are only $5$ and $12$ multipoles
having a probability of $P \leq 5\%$  compared to that of simulations for COMMANDER and NILC
maps respectively.
Similarly, by filling the masked region with the individual masks available with  each
component separation map (PMCMDR and PMNILC), we find that there are $7$ and $12$ multipoles
with $P \le 0.05$ for COMMANDER and NILC $E$-mode polarization maps respectively.
Thus applying various masks reveals the (in)stability of anomalous multipoles against
galactic cuts and potential foreground contamination.
The corresponding cumulative probabilities in various cases are listed in the \emph{third} column
of Table~(\ref{tab:tab2}).

\begin{table}
\scriptsize
\centering
\begin{tabular}{c l c}
\hline
\hline
Map/Mask & Multipoles & Cumulative \\
         &            & Probability \\
\hline
\hline
 & & \\
COMMANDER/ & 58, 76, 98, 108, 146 & $ 0.656 $ \\
   UPB77     &                      & \\
 & & \\
COMMANDER/ & 57, 58, 76, 95, 98, 108, & $ 0.320 $ \\
  PMCMDR     & 146             & \\
 & & \\
NILC/     & 45, 51, 54, 56, 60, 78,    & $9.726 \times 10^{-3}$ \\
UPB77       & 81, 84, 92, 95, 98, 108    & \\
 & & \\
NILC/     & 56, 60, 66, 78, 84, 90,    & $9.726 \times 10^{-3}$ \\
PMNILC      & 92, 95, 104, 108, 122,     & \\
            & 149                        & \\
 & & \\
\hline
\hline
\end{tabular}
\caption{ Same as Table~(\ref{tab:tab1}), but for full sky $E$-polarization maps constructed by filling
          the potentially contaminated galactic region as defined by various masks (see section~\ref{sec:data}
          for more details).
          The two component separated polarization maps considered here, were studied using both the
          individual polarization masks - the PMCMDR and PMNILC masks, as well as the common UPB77 mask
          shown in Fig.~(\ref{fig:mask-pol}).
          The \emph{second} column lists the anomalous multipoles that are outside the $2\sigma$ confidence
          level, and the \emph{third} column lists the cumulative probabilities for finding the observed
          number or more of the statistically anisotropic multipoles in the range $l=40-150$, that have
          a $P$-value $\leq 5\%$.}
\label{tab:tab2}
\end{table}

From Table~(\ref{tab:tab2}), one notices that the cumulative probabilities are larger in the
\emph{filled} sky $E$-mode polarization maps than cleaned \emph{full} sky maps, evidently
owing to the decrease in number of anomalous multipoles in the later case.

\begin{figure*}
    \centering
\includegraphics[width=0.95\textwidth]{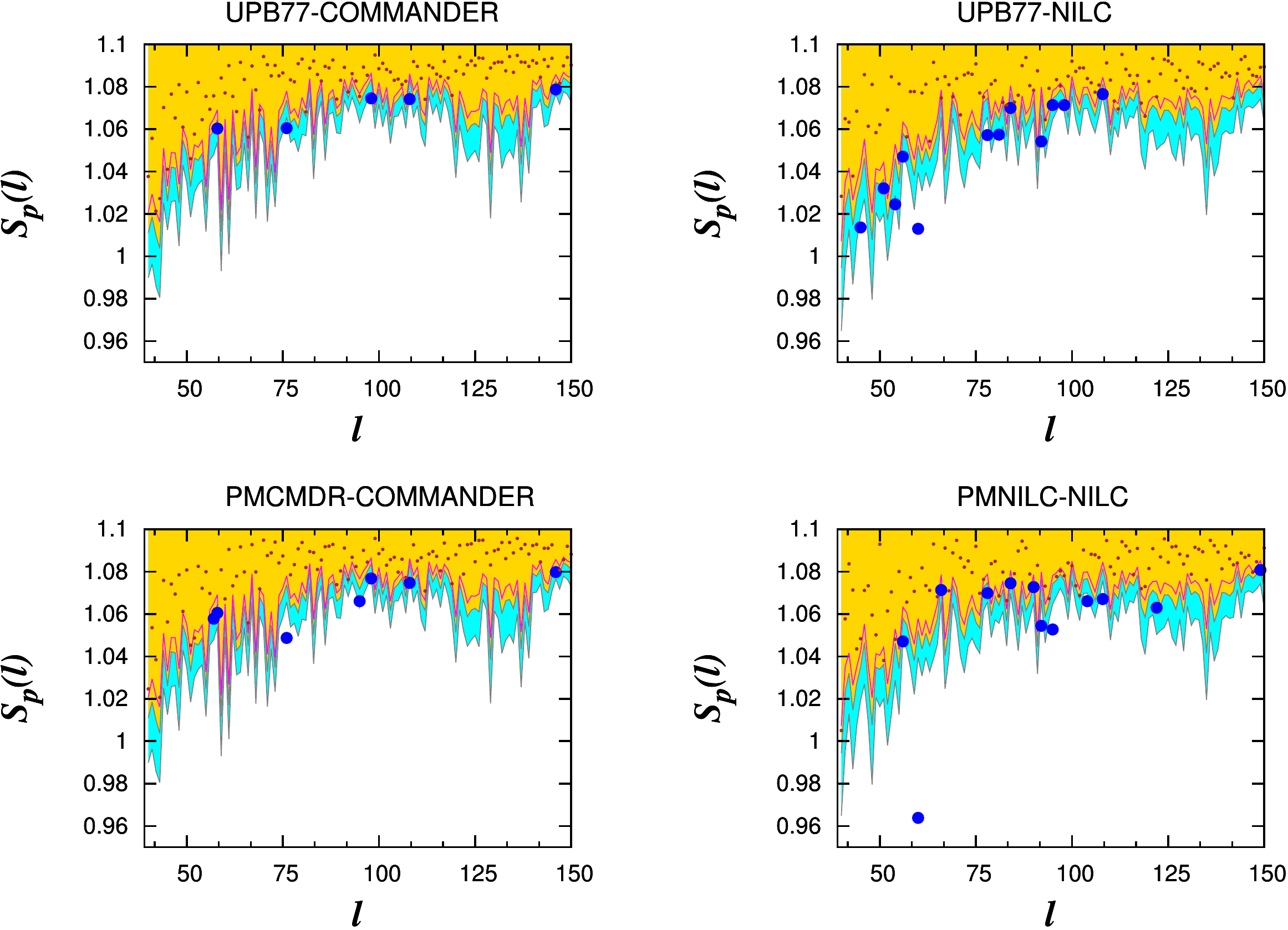}
    \caption{Plotted here are the Power entropy ($S_p$) values as obtained from full sky
             PLANCK's COMMANDER and NILC $E$-maps constructed after filling the masked
             sky defined by various polarization masks shown in Fig.~(\ref{fig:mask-pol}).
             The \emph{first} and \emph{second} columns correspond to the Power entropy
             for the range $l=40-150$ as obtained from COMMANDER and NILC $E$-polarization
             maps. The \emph{top} and \emph{bottom} figures in each column correspond to
             the use of UPB77 common polarization mask and the component separation
             specific PMCMDR/PMNILC mask on respective cleaned maps.
             The color coding of data points and confidence contours is same as in
             Fig.~(\ref{fig:counterplot1}).}
    \label{fig:counterplot2}
\end{figure*}

In Fig.~(\ref{fig:counterplot2}), we show the Power entropy values from data in comparison
to those derived from simulations. As mentioned earlier in section~\ref{sec:data}, the observed
$S_p$ values for each multipole are obtained as average value of the statistic over $100$ random
fillings of the data to construct full sky $E$-mode polarization maps.
The contours of $90\%$, $95\%$ and $99\%$ confidence levels from simulations are shown as
\emph{magenta} line, \emph{gold} band, and \emph{cyan} band respectively. The data points
are shown as \emph{red} dots, while those which are outside the $2\sigma$ contour are denoted by
\emph{blue} points. This contour plot neatly highlights various multipoles which are inconsistent
with the isotropic predictions.
So, we may now say that presence of residual foregrounds indeed had an effect on our isotropy
test when full sky polarization maps are used as provided.
In both COMMANDER and NILC maps, the number of anomalous multipoles nearly reduced by half
when full sky maps are constructed using UPB77 mask. However since noise is dominant in the
PLANCK polarization maps, the stability of these modes can only be validated in the future.

\subsection{Alignments across multipoles}
\label{subsec:AlignEntropy}

In this section, we discuss alignments among mutlipoles in the chosen multipole range
 using Alignment entropy, $S_X$, as defined in Eq.~(\ref{eq:alignent}). 
We divide the chosen range $l=40-150$ into $10$ multipole bins having $11$ multipoles
per bin. Thus the multipole bins we analyze are $l=40-50$, $51-61$, $62-72$, $73-83$,
$84-94$, $95-105$, $106-116$, $117-127$, $128-138$ and $139-149$.
The statistic, $S_X$, is computed for each multipole bin and its significance is
estimated using simulations that are generated as discussed in section~\ref{sec:data}.

\subsubsection{Full sky analysis}
Here we probe for any coherent alignments across multipoles in $E$-mode polarization
maps as obtained using PLANCK's full sky Stokes $Q/U$ CMB maps, derived using COMMANDER
and NILC cleaning procedures. In each multipole bin of the data maps, Alignment entropy
is calculated. The statistical significance of $S_X$ are obtained by comparing the
data statistic with $4000$ isotropic random CMB $E$-mode filtered polarization maps
added with the noise proxy (HRHD map) of respective component separation methods.

\begin{figure}
\centering
\includegraphics[width=0.48\textwidth]{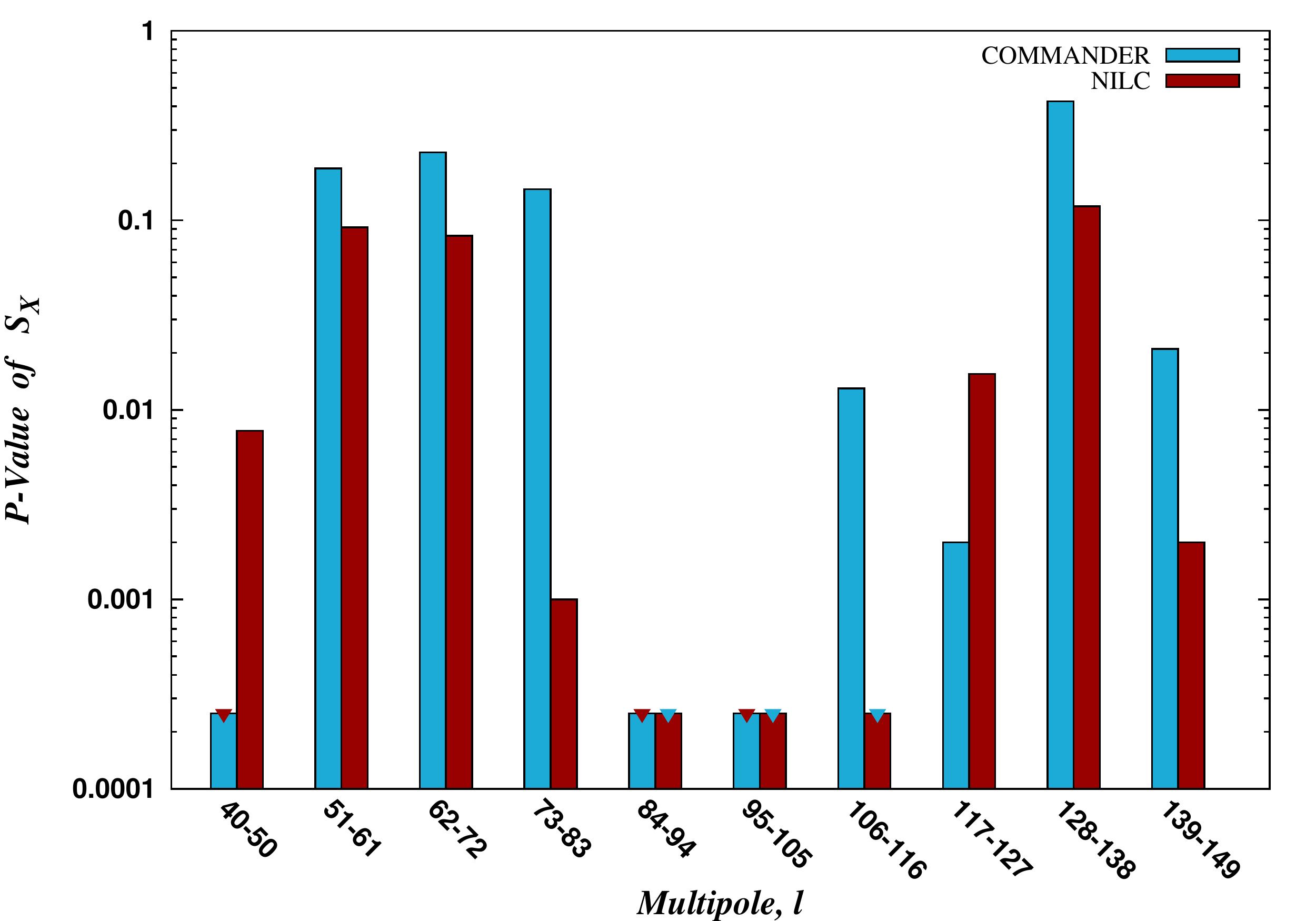}
\caption{$P$-values of Alignment entropy, $S_X$, corresponding to various multipole bins in the
         range $l=40-150$ (divided into 10 bin) are plotted here.
         The two bars at each mutlipole bin as indicated correspond to the significances
         of Alignment entropy for that bin from CMB $E$-mode polarization maps derived
         from full sky Stokes $Q/U$ COMMANDER and NILC maps.
         Those histogram bars which have a triangle at the top, denote a significance
         $<1/4000$.}
\label{fig:algn-entrp-fullsky-E}
\end{figure}

The list of multipole bins and the significance of $S_X$ for these bins are listed
in \emph{second} and \emph{third} column of Table~(\ref{tab:tab5}) respectively.
As we can see from that table,
the multipole bins $40-50$, $84-94$, $95-105$, $106-116$, $117-127$, and $139-149$
have a $P$-value $\le 0.05$ for Alignment entropy. We extract the common alignment
vector from PEVs for a bundle of multipoles using the Alignment tensor, $X$, defined in
Eq.~(\ref{eq:alignten}). The common alignment vector is taken as the eigenvector
corresponding to the largest eigenvalue of the Alignment tensor matrix.
The common alignment vector direction of those bins in galactic co-ordinates, $(l,b)$,
having $P$-value $\leq 5\%$ for $S_X$ are also listed in Table~(\ref{tab:tab5}),
in the \emph{fourth} column.
We notice that these vectors almost lie in the  galactic region. The effect of
the presence of residual contamination on these vectors will be assessed in the
next section.
For ease of comparison, we also show the observed significances of $S_X$
in various multipole bins from full sky COMMANDER and NILC $E$-maps in
Fig.~(\ref{fig:algn-entrp-fullsky-E}).
The Alignment entropy from data corresponding to some of the multipole bins was always
smaller than the simulations. Therefore we denoted those histogram bars with a triangle
at the top to indicate that the significance of data statistic is $<1/4000$.

\begin{table}
\scriptsize
\centering
\begin{tabular}{|c c c c}
\hline
\hline
Map & Multipole Range & $P$-value & $(l,b)$ \\
\hline
\hline
COMMANDER         & $40-50$   & $<1/4000$ & $(87^\circ,27^\circ)$\\
                  & $51-61$   & $0.19$    & - \\
                  & $62-72$   & $0.23$    & - \\
                  & $73-83$   & $0.15$    & - \\
                  & $84-94$   & $<1/4000$ & $(95^\circ,19^\circ)$\\
                  & $95-105$  & $<1/4000$ & $(96^\circ,30^\circ)$\\
                  & $106-116$ & $0.013$   & $(75^\circ,18^\circ)$ \\
                  & $117-127$ & $0.002$   & $(103^\circ,33^\circ)$\\
                  & $128-138$ & $0.424$   & -\\
                  & $139-149$ & $0.021$   & $(90^\circ,26^\circ)$\\
\hline
NILC              & $40-50$   & $0.008$   & $(87^\circ,24^\circ)$\\
                  & $51-61$   & $0.092$   & -\\
                  & $62-72$   & $0.083$   & -\\
                  & $73-83$   & $0.001$   & $(99^\circ,29^\circ)$\\
                  & $84-94$   & $<1/4000$ & $(93^\circ,31^\circ)$\\
                  & $95-105$  & $<1/4000$ & $(95^\circ,29^\circ)$\\
                  & $106-116$ & $<1/4000$ & $(82^\circ,21^\circ)$\\
                  & $117-127$ & $0.02$    & $(96^\circ,36^\circ)$\\
                  & $128-138$ & $0.12$    & -\\
                  & $139-149$ & $0.002$   & $(98^\circ,46^\circ)$\\
\hline
\hline
\end{tabular}
\caption{Significance of Alignment entropy ($S_X$) and direction of common alignment axis
          obtained using Alignment tensor ($X$) corresponding to various multipole bins
         of the cleaned $E$-mode polarization maps, derived from full sky Stokes $Q/U$
         COMMANDER and NILC maps, are listed here.
         The range $l=40-150$ is divided into $10$ multipole bins. The $P$-value of $S_X$
         for various bins are given in the \emph{third} column.
         Where none of the simulations yield a value less than the data statistic,
         it's significance is listed as $<1/4000$.
         The direction of the common alignment axis for those bins
         whose $P$-value is $\leq 0.05$ are given in the \emph{fourth} column.}
\label{tab:tab5}
\end{table}

\subsubsection{Filled sky analysis}
In order to understand the likely effect of residual contamination in the PLANCK full sky
CMB polarization maps on our collective alignments' study, here we use filled $E$-mode
data polarization map, generated as discussed in section~\ref{sec:data}.
The common $UPB77$ mask is first applied on the cleaned PLANCK's COMMANDER and NILC $IQU$
maps and  the statistic $S_X$ is obtained as mean value of the same quantity from filled
$IQU$ maps constructed using 100 random realizations of CMB with HRHD noise proxy
to fill the masked portion in the data maps.
The filling is also performed using the other two polarization masks shown in Fig.~(\ref{fig:mask-pol}),
viz., the component specific $PMCMDR$ and $PMNILC$ masks corresponding to COMMANDER and NILC
foreground cleaning schemes.
The statistical significance of the observed value of Alignment entropy is studied using
$4000$ CMB $E$-mode realizations with noise.

\begin{table}
\scriptsize
\centering
\begin{tabular}{c c c c}
\hline
\hline
Map/Mask & Multipole bin & $P$-value & $(l,b)$  \\
\hline
\hline
COMMANDER/ &  $40-50$   & $0.0005$ & $(85^\circ,24^\circ)$  \\
 UPB77     &  $51-61$   & $0.106$  & - \\
           &  $62-72$   & $0.273$  & - \\
           &  $73-83$   & $0.056$  & - \\
           &  $84-94$   & $0.002$  & $(91^\circ,22^\circ)$   \\
           &  $95-105$  & $0.025$  & $(98^\circ,47^\circ)$   \\
           &  $106-116$ & $0.014$  & $(72^\circ,29^\circ)$ \\
           &  $117-127$ & $0.064$  & - \\
           &  $128-138$ & $0.123$  & - \\
           &  $139-149$ & $0.012$  & $(88^\circ,34^\circ)$\\
\hline
COMMANDER/ &  $40-50$   & $0.0005$ & $(85^\circ,31^\circ)$\\
 PMCMDR    &  $51-61$   & $0.055$  & - \\
           &  $62-72$   & $0.13$   & - \\
           &  $73-83$   & $0.58$   & - \\
           &  $84-94$   & $0.0013$ & $(89^\circ,27^\circ)$ \\
           &  $95-105$  & $0.007$  & $(98^\circ,40^\circ)$ \\
           &  $106-116$ & $0.0008$ & $(72^\circ,28^\circ)$ \\
           &  $117-127$ & $0.05$   & - \\
           &  $128-138$ & $0.156$  & - \\
           &  $139-149$ & $0.0234$ & $(91^\circ,33^\circ)$\\
\hline
NILC/     &  $40-50$   & $<1/4000$ & $(93^\circ,26^\circ)$ \\
UPB77     &  $51-61$   & $0.007$   & $(101^\circ,35^\circ)$\\
          &  $62-72$   & $0.342$   & - \\
          &  $73-83$   & $0.137$   & - \\
          &  $84-94$   & $<1/4000$ & $(92^\circ,16^\circ)$ \\
          &  $95-105$  & $0.0003$  & $(100^\circ,36^\circ)$ \\
          &  $106-116$ & $<1/4000$ & $(84^\circ,27^\circ)$ \\
          &  $117-127$ & $0.002$   & $(99^\circ,25^\circ)$ \\
          &  $128-138$ & $0.002$   & $(92^\circ,38^\circ)$ \\
          &  $139-149$ & $0.002$   & $(90^\circ,34^\circ)$ \\
\hline
NILC/     &  $40-50$   & $0.003$   & $(90^\circ,26^\circ)$\\
PMNILC    &  $51-61$   & $0.002$   & $(102^\circ,38^\circ)$ \\
          &  $62-72$   & $0.47$    & - \\
          &  $73-83$   & $0.03$    & $(105^\circ,28^\circ)$\\
          &  $84-94$   & $<1/4000$ & $(97^\circ,29^\circ)$ \\
          &  $95-105$  & $<1/4000$ & $(100^\circ,31^\circ)$ \\
          &  $106-116$ & $<1/4000$ & $(80^\circ,20^\circ)$ \\
          &  $117-127$ & $0.001$   & $(98^\circ,27^\circ)$ \\
          &  $128-138$ & $0.08$    & - \\
          &  $139-149$ & $0.002$   & $(99^\circ,43^\circ)$ \\
\hline
\hline
\end{tabular}
\caption{Same as Table~(\ref{tab:tab5}), but the significances of $S_X$ and direction
         of common alignment axis for various bins correspond to filled sky polarization
         maps as indicated (see text for detials).
         The $P$-values are listed in \emph{third} column and the common alignment vectors
         for only those bins whose Alignment entropy, $S_X$, is found to be anomalous at
         $2\sigma$ level are listed in \emph{fourth} column.}
\label{tab:tab7}
\end{table}

The list of multipole bins and the
significance of $S_X$ from these bins are given in Table~(\ref{tab:tab7}). Now, we find
that the multipole bins, $l=40-50$, $84-94$, $95-105$, $106-116$, and $139-149$ have
$P\leq 0.05$ using the UPB77 mask, and $l=40-50$, $84-94$, $95-105$, $106-116$, $117-127$,
and $139-149$ multipole bins have $P$-value $\leq 0.05$ with PMCMDR mask on COMMANDER
polarization map. Similarly for NILC polarization map, the multipole bins that are anomalous
at $2\sigma$ level are $l=40-50$, $51-61$, $84-94$, $95-105$, $106-116$, $117-127$, $128-138$,
and $139-149$ when UPB77 mask was used, and $l=40-50$, $51-61$, $73-83$, $84-94$, $95-105$,
$106-116$, $117-127$, $128-138$, and $139-149$ bins when PMNILC mask was used.

\begin{figure}
\centering
\includegraphics[width=0.48\textwidth]{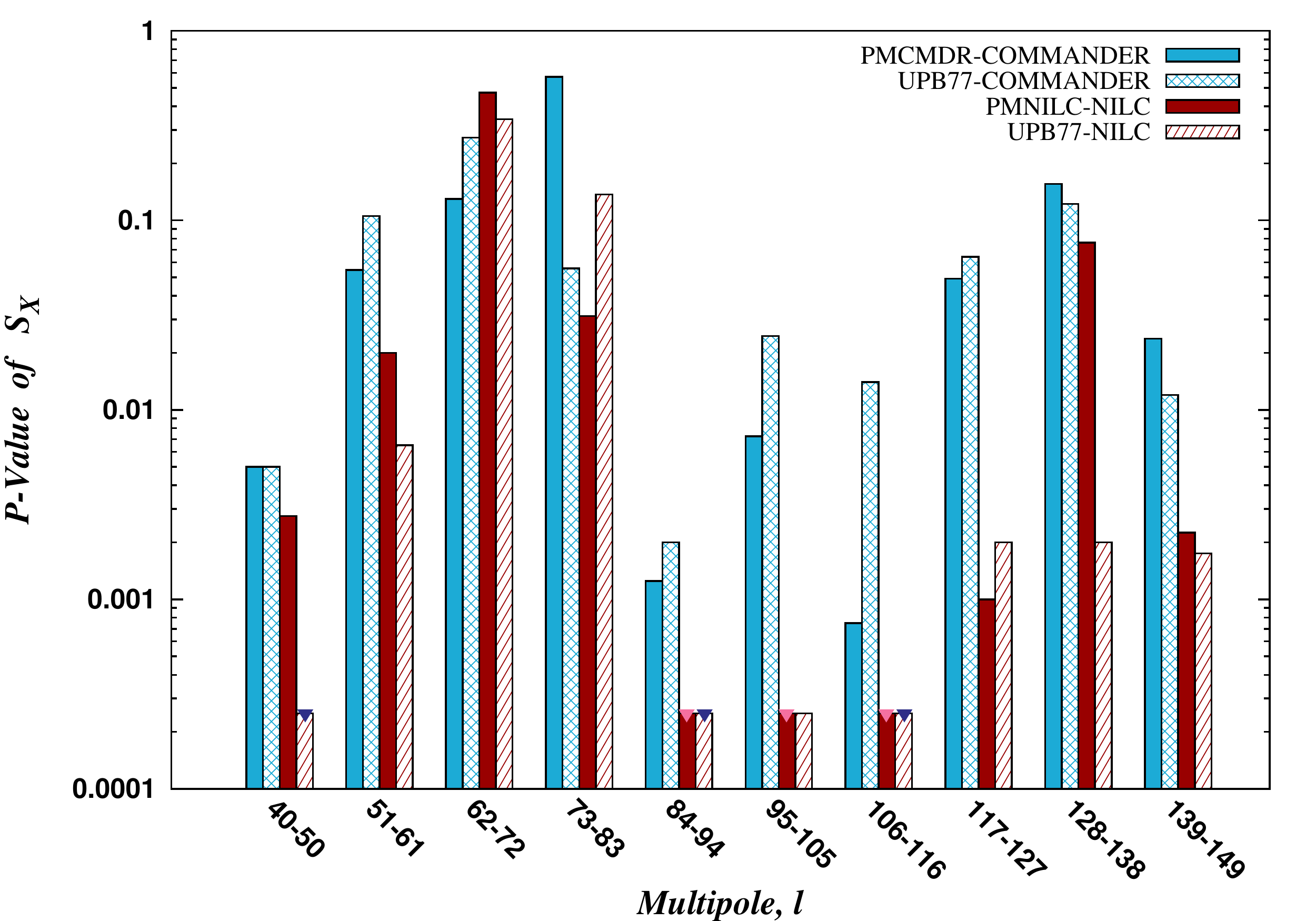}
\caption{Same as Fig.~(\ref{fig:algn-entrp-fullsky-E}), but for the filled sky
        CMB polarization maps viz, COMMANDER $E$-map by filling the masked region
        defined by the common UPB77 and PMCMDR masks, and the NILC $E$-map where
        the inside of the masking portion defined by UPB77 and PMNILC
        masks is filled, using  a random CMB realization and HRHD noise proxy.
        The data statistic is computed as average of the same quantity extracted
        from 100 such random fillings of the data, which is then used to compute significances.
        Here also, those multipole bins for a given component separation map/mask combination
        where the significance was found to be $<1/4000$, their histogram bar is plotted with
        a triangle at the top.}
\label{fig:algn-entrp-filledky-E}
\end{figure}

The collective alignment vector direction for various bins obtained using Alignment tensor, $X$,
(Eq.~(\ref{eq:alignten})) which are found to have a $P$-value $\leq 5\%$ for $S_X$ are
tabulated in the last column of Table~(\ref{tab:tab7}). Even in this case, we see that
these collective alignment vectors lie closer to the galactic region. Hence these modes
found to be anomalous may still be affected by residual foreground bias.
In Fig.~(\ref{fig:algn-entrp-filledky-E}), the $P$-values of $S_X$ found in the filled
sky case for various multipole bins and masks used are shown as histograms for ready comparison.

\section{Conclusion}
\label{sec:con}
In the present work, we studied violation of isotropy of various modes that are available
in the cleaned CMB polarization maps from PLANCK full mission data release. Specifically
we scrutinized the multipole range $l=40-150$ of the $E$-mode CMB maps derived using the
COMMANDER and NILC $Q/U$ polarization maps. We applied our symmetry based \emph{Power tensor}
method to test statistical (an)isotropy of these maps.
A set of 4000 simulations were generated using the theoretical angular power spectrum
obtained using the best fit cosmological parameters from PLANCK 2015 release in the CAMB
software package. These realizations were smoothed with appropriate beam window function,
and filtered using a high-pass cosine filter to retain only those modes in simulations,
that are currently made available in the data.
The half-ring half-difference (HRHD) maps of respective component separation methods were
taken as noise proxy, and added with the smoothed, filtered $E$-mode random CMB realizations
to generate mock observed maps.

In order to understand the effect of potential bias due to residual foregrounds in the full sky
cleaned polarization maps on our isotropy studies, we generated another set of full sky
maps where we filled part of the sky that is omitted by the UPB77 polarization mask with
100 appropriate random CMB maps with noise. All the statistics computed in the filled sky
case are taken as average value of the same quantity over these 100 \emph{filled} full sky
$E$-mode data maps. We also used the individual polarization masks from COMMANDER and NILC
foreground removal methods. We reiterate that the filling procedure we employed is only expected
to lower any signal of statistical anisotropy if present, because of the randomizing nature
of the filling process.

We note the following observations. The number of anomalous multipoles that indicate isotropy
violation at the level of $2\sigma$ in the full sky COMMNADER and NILC polarization maps
are found to be $9$ and $21$ respectively. However when this same range $l=40-150$ is analyzed
using filled sky maps constructed using the conservative UPB77 polarization mask, the number of
anomalous modes have reduced to nearly half. The number remains albeit the same when component
specific polarization masks are used. Thus we may say that the galactic residuals indeed have an
effect on our test of isotropy of various multipoles. It is interesting to note that the
number of anomalous multipoles with $P$-value $\leq 0.05$ are more in NILC CMB $E$-map than
in the COMMANDER map. This observation is particularly interesting given that the recovered
CMB signal using NILC procedure is supposed to be very reliable over a much larger
fraction of the sky than the COMMANDER map.
The respective masking fractions of COMMANDER and NILC polarization masks are $\approx 83\%$
and $96\%$. The effective probability of finding the observed number of statistically anisotropic
modes in the range $l=40-150$ in NILC map using various masks is correspondingly low.

We then studied alignments across multipoles using \emph{Alignment entropy} over the
chosen range $l=40-150$, divided into 10 blocks with 11 multipole per bin.
Here also we analyzed full sky as well as filled sky maps for coherent alignments
across multipoles using 4000 simulations. All three polarization masks considered
in the preceding analysis were applied to data $E$-maps to understand any foreground biases.
We find a tentative evidence for collective alignment in some of the multipole blocks.
However, we also found that the direction of these common alignment axes lie closer to the
galactic plane in both full sky and filled sky cases. Hence the effect of galactic bias,
even after filling the potentially contaminated regions, may still be significant. In these
alignment tests as well, we find that more number of multipole bins are anomalous in NILC map
compared to COMMANDER CMB $E$-map.

Thus, the modes currently available in the polarization maps we analyzed appear to be sensitive
to various galactic cuts. Further, the number and actual modes that are anomalous change, by applying
different masks with few common multipoles surviving among them. These
multipoles may also be effected by noise, given that the noise levels are significant in PLANCK
polarization measurements. In light of these, an analysis of improved polarization maps from PLANCK
in a future data release can only confirm our findings. In this work, however, we didn't
make any attempt to identify the exact cause of the observed isotropy violation in polarization maps.

\section*{Acknowledgements}
We thank the anonymous referee for a careful reading of our manuscript and
the suggestions made, which greatly helped in improving clarity and
overall presentation of our work.
Some of the results in the current work were derived using the publicly available
HEALPix package\footnote{\url{https://healpix.jpl.nasa.gov/}}.
We also acknowledge the use of CAMB\footnote{\url{http://camb.info/}}, a freely
available Boltzmann solver for CMB anisotropies.
Part of the results presented here are based on observations obtained with
PLANCK\footnote{\url{http://www.esa.int/Planck}}, an ESA science mission with instruments
and contributions directly funded by ESA Member States, NASA, and Canada.


\bsp	

\label{lastpage}


\begin{thebibliography}{99}
\makeatletter
\relax
\def\mn@urlcharsother{\let\do\@makeother \do\$\do\&\do\#\do\^\do\_\do\%\do\~}
\def\mn@doi{\begingroup\mn@urlcharsother \@ifnextchar [ {\mn@doi@}
  {\mn@doi@[]}}
\def\mn@doi@[#1]#2{\def\@tempa{#1}\ifx\@tempa\@empty \href
  {http://dx.doi.org/#2} {doi:#2}\else \href {http://dx.doi.org/#2} {#1}\fi
  \endgroup}
\def\mn@eprint#1#2{\mn@eprint@#1:#2::\@nil}
\def\mn@eprint@arXiv#1{\href {http://arxiv.org/abs/#1} {{\tt arXiv:#1}}}
\def\mn@eprint@dblp#1{\href {http://dblp.uni-trier.de/rec/bibtex/#1.xml}
  {dblp:#1}}
\def\mn@eprint@#1:#2:#3:#4\@nil{\def\@tempa {#1}\def\@tempb {#2}\def\@tempc
  {#3}\ifx \@tempc \@empty \let \@tempc \@tempb \let \@tempb \@tempa \fi \ifx
  \@tempb \@empty \def\@tempb {arXiv}\fi \@ifundefined
  {mn@eprint@\@tempb}{\@tempb:\@tempc}{\expandafter \expandafter \csname
  mn@eprint@\@tempb\endcsname \expandafter{\@tempc}}}

\bibitem[\protect\citeauthoryear{{Aiola}, {Wang}, {Kosowsky}, {Kahniashvili} \& 
	{Firouzjahi}}{{Aiola}  et~al.}{2015}]{Aiola2015}
{Aiola} S., {Wang} B., {Kosowsky} A., {Kahniashvili} T.,
	{Firouzjahi} H.,  2015, \mn@doi [Phys. Rev.
  D] {10.1103/PhysRevD.92.063008}, \href
  {http://adsabs.harvard.edu/abs/2015PhRvD..92f3008A} {92, 063008}

\bibitem[\protect\citeauthoryear{{Akrami}, {Fantaye}, {Shafieloo}, {Eriksen},
  {Hansen}, {Banday}  \& {G{\'o}rski}}{{Akrami} et~al.}{2014}]{Akrami2014}
{Akrami} Y.,  {Fantaye} Y.,  {Shafieloo} A.,  {Eriksen} H.~K.,  {Hansen} F.~K.,
   {Banday} A.~J.,   {G{\'o}rski} K.~M.,  2014, \mn@doi [Astrophysical
  Journal Letters] {10.1088/2041-8205/784/2/L42}, \href
  {http://adsabs.harvard.edu/abs/2014ApJ...784L..42A} {784, L42}

\bibitem[\protect\citeauthoryear{{Aluri} \& {Jain}}{{Aluri} \&
  {Jain}}{2012}]{Aluri2012}
{Aluri} P.~K.,  {Jain} P.,  2012, \mn@doi [Mon. Not. R. Astron. Soc.]
  {10.1111/j.1365-2966.2011.19981.x}, \href
  {http://adsabs.harvard.edu/abs/2012MNRAS.419.3378A} {419, 3378}

\bibitem[\protect\citeauthoryear{{Aluri} \& {Rath}}{{Aluri} \&
  {Rath}}{2016}]{Aluri2016}
{Aluri} P.~K.,  {Rath} P.~K.,  2016, \mn@doi [Mon. Not. R. Astron. Soc.]
  {10.1093/mnras/stw283}, \href
  {http://adsabs.harvard.edu/abs/2016MNRAS.458.4269A} {458, 4269}

\bibitem[\protect\citeauthoryear{{Aluri}, {Ralston}  \& {Weltman}}{{Aluri}
  et~al.}{2017}]{Aluri2017}
{Aluri} P.~K.,  {Ralston} J.~P.,  {Weltman} A.,  2017, \mn@doi [Mon. Not. R. Astron. Soc.]
  {10.1093/mnras/stx2112}, \href
  {http://adsabs.harvard.edu/abs/2017MNRAS.472.2410A} {472, 2410}

\bibitem[\protect\citeauthoryear{{Bennett}, {Hill}, {Hinshaw},et~al.,}{{Bennett}
 et~al.}{2011}]{wmap7yranom}
 {Bennett} C.~L., {Hill} R.~S., {Hinshaw} G., et~al., 2011,
  \mn@doi [Astrophysical Journal]
  {10.1088/0067-0049/192/2/17}, \href
  {http://adsabs.harvard.edu/abs/2011ApJS..192...17B} {192, 17}

\bibitem[\protect\citeauthoryear{{Bielewicz}, {Eriksen}, {Banday}, {Gorski}  \&
  {Lilje}}{{Bielewicz} et~al.}{2005}]{Bielewicz2005}
{Bielewicz} P.,  {Eriksen} H.~K.,  {Banday} A.~J.,  {Gorski} K.~M.,   {Lilje}
  P.~B.,  2005, \mn@doi [Astrophysical
  Journal] {10.1086/497263}, \href
  {http://adsabs.harvard.edu/abs/2005ApJ...635..750B} {635, 750}

\bibitem[\protect\citeauthoryear{{Birch}}{{Birch}}{1982}]{Birch1982}
{Birch} P.,  1982, \mn@doi [Nature] {10.1038/298451a0}, \href
  {http://adsabs.harvard.edu/abs/1982Natur.298..451B} {298, 451}

\bibitem[\protect\citeauthoryear{{Copi}, {Huterer}  \& {Starkman}}{{Copi}
  et~al.}{2004}]{Copi2004}
{Copi} C.~J.,  {Huterer} D.,   {Starkman} G.~D.,  2004, \mn@doi [Phys. Rev. D]
  {10.1103/PhysRevD.70.043515}, \href
  {http://adsabs.harvard.edu/abs/2004PhRvD..70d3515C} {70, 043515}

\bibitem[\protect\citeauthoryear{{Copi}, {Huterer}, {Schwarz}  \&
  {Starkman}}{{Copi} et~al.}{2006}]{Copi2006}
{Copi} C.~J.,  {Huterer} D.,  {Schwarz} D.~J.,   {Starkman} G.~D.,  2006,
  \mn@doi [Mon. Not. R. Astron. Soc.] {10.1111/j.1365-2966.2005.09980.x}, \href
  {http://adsabs.harvard.edu/abs/2006MNRAS.367...79C} {367, 79}

\bibitem[\protect\citeauthoryear{{Cruz}, {Mart{\'{\i}}nez-Gonz{\'a}lez},
  {Vielva}  \& {Cay{\'o}n}}{{Cruz} et~al.}{2005}]{Cruz2005}
{Cruz} M.,  {Mart{\'{\i}}nez-Gonz{\'a}lez} E.,  {Vielva} P.,   {Cay{\'o}n} L.,
  2005, \mn@doi [Mon. Not. R. Astron. Soc.] {10.1111/j.1365-2966.2004.08419.x},
  \href {http://adsabs.harvard.edu/abs/2005MNRAS.356...29C} {356, 29}

\bibitem[\protect\citeauthoryear{{Cruz}, {Tucci},
  {Mart{\'{\i}}nez-Gonz{\'a}lez}  \& {Vielva}}{{Cruz} et~al.}{2006}]{Cruz2006}
{Cruz} M.,  {Tucci} M.,  {Mart{\'{\i}}nez-Gonz{\'a}lez} E.,   {Vielva} P.,
  2006, \mn@doi [Mon. Not. R. Astron. Soc.] {10.1111/j.1365-2966.2006.10312.x},
  \href {http://adsabs.harvard.edu/abs/2006MNRAS.369...57C} {369, 57}

\bibitem[\protect\citeauthoryear{{Cruz}, {Mart{\'{\i}}nez-Gonz{\'a}lez},
  {Vielva}, {Diego}, {Hobson}  \& {Turok}}{{Cruz} et~al.}{2008}]{Cruz2008}
{Cruz} M.,  {Mart{\'{\i}}nez-Gonz{\'a}lez} E.,  {Vielva} P.,  {Diego} J.~M.,
  {Hobson} M.,   {Turok} N.,  2008, \mn@doi [Mon. Not. R. Astron. Soc.]
  {10.1111/j.1365-2966.2008.13812.x}, \href
  {http://adsabs.harvard.edu/abs/2008MNRAS.390..913C} {390, 913}

\bibitem[\protect\citeauthoryear{{de Oliveira-Costa}, {Tegmark}, {Zaldarriaga}
  \& {Hamilton}}{{de Oliveira-Costa} et~al.}{2004}]{Costa2004}
{de Oliveira-Costa} A.,  {Tegmark} M.,  {Zaldarriaga} M.,   {Hamilton} A.,
  2004, \mn@doi [Phys. Rev. D] {10.1103/PhysRevD.69.063516}, \href
  {http://adsabs.harvard.edu/abs/2004PhRvD..69f3516D} {69, 063516}

\bibitem[\protect\citeauthoryear{{Eriksen}, {Hansen}, {Banday}, {Gorski}  \&
  {Lilje}}{{Eriksen} et~al.}{2004}]{Eriksen2004a}
{Eriksen} H.~K.,  {Hansen} F.~K.,  {Banday} A.~J.,  {Gorski} K.~M.,   {Lilje}
  P.~B.,  2004, \mn@doi [Astrophysical
  Journal] {10.1086/382267}, \href
  {http://adsabs.harvard.edu/abs/2004ApJ...605...14E} {605, 14}

\bibitem[\protect\citeauthoryear{{Gorski}, {Hivon}, {Banday}, {Wandelt},
  {Hansen}, {Reinecke}  \& {Bartelmann}}{{Gorski} et~al.}{2005}]{Gorski2005}
{Gorski} K.,  {Hivon} E.,  {Banday} A.~J.,  {Wandelt} B.~D.,  {Hansen} F.~K.,
  {Reinecke} M.,   {Bartelmann} M.,  2005, \mn@doi [Astrophysical Journal]
  {10.1086/427976}, \href {http://adsabs.harvard.edu/abs/2005ApJ...622..759G}
  {622, 759}

\bibitem[\protect\citeauthoryear{{Hajian} \& {Souradeep}}{{Hajian} \&
  {Souradeep}}{2003}]{Hajian2003}
{Hajian} A.,  {Souradeep} T.,  2003, \mn@doi [Astrophysical
  Journal] {10.1086/379757},
  \href {http://adsabs.harvard.edu/abs/2003ApJ...597L...5H} {597, L5}

\bibitem[\protect\citeauthoryear{{Hajian} \& {Souradeep}}{{Hajian} \&
  {Souradeep}}{2006}]{Hajian2006}
{Hajian} A.,  {Souradeep} T.,  2006, \mn@doi [Phys. Rev. D]
  {10.1103/PhysRevD.74.123521}, \href
  {http://adsabs.harvard.edu/abs/2006PhRvD..74l3521H} {74, 123521}

\bibitem[\protect\citeauthoryear{{Hansen}, {Banday}, {Gorski}, {Eriksen}  \&
  {Lilje}}{{Hansen} et~al.}{2009}]{Hansen2009}
{Hansen} F.~K.,  {Banday} A.~J.,  {Gorski} K.~M.,  {Eriksen} H.~K.,   {Lilje}
  P.~B.,  2009, \mn@doi [Astrophysical Journal] {10.1088/0004-637X/704/2/1448},
  \href {http://adsabs.harvard.edu/abs/2009ApJ...704.1448H} {704, 1448}

\bibitem[\protect\citeauthoryear{{Hanson} \& {Lewis}}{{Hanson} \&
  {Lewis}}{2009}]{Hanson2009}
{Hanson} D.,  {Lewis} A.,  2009, \mn@doi [Phys. Rev. D]
  {10.1103/PhysRevD.80.063004}, \href
  {http://adsabs.harvard.edu/abs/2009PhRvD..80f3004H} {80, 063004}

\bibitem[\protect\citeauthoryear{{Hoftuft}, {Eriksen}, {Banday}, {Gorski},
  {Hansen}  \& {Lilje}}{{Hoftuft} et~al.}{2009}]{Hoftuft2009}
{Hoftuft} J.,  {Eriksen} H.~K.,  {Banday} A.~J.,  {Gorski} K.~M.,  {Hansen}
  F.~K.,   {Lilje} P.~B.,  2009, \mn@doi [Astrophysical
  Journal]
  {10.1088/0004-637X/699/2/985}, \href
  {http://adsabs.harvard.edu/abs/2009ApJ...699..985H} {699, 985}

\bibitem[\protect\citeauthoryear{{Howlett}, {Lewis}, {Hall}  \&
  {Challinor}}{{Howlett} et~al.}{2012}]{Howlett2012}
{Howlett} C.,  {Lewis} A.,  {Hall} A.,   {Challinor} A.,  2012, \mn@doi
  [Journal of Cosmology and Astroparticle Physics]
  {10.1088/1475-7516/2012/04/027}, \href
  {http://adsabs.harvard.edu/abs/2012JCAP...04..027H} {4, 027}

\bibitem[\protect\citeauthoryear{{Hutsemekers}}{{Hutsemekers}}{1998}]{Hutsemekers1998}
{Hutsemekers} D.,  1998, Astronomy and Astrophysics, \href
  {http://adsabs.harvard.edu/abs/1998A%26A...332..410H} {332, 410}

\bibitem[\protect\citeauthoryear{{Hutsemekers} \& {Lamy}}{{Hutsemekers} \&
  {Lamy}}{2001}]{Hutsemekers2001}
{Hutsemekers} D.,  {Lamy} H.,  2001, \mn@doi [Astronomy and Astrophysics]
  {10.1051/0004-6361:20000443}, \href
  {http://adsabs.harvard.edu/abs/2001A%26A...367..381H} {367, 381}

\bibitem[\protect\citeauthoryear{{Jain} \& {Ralston}}{{Jain} \&
  {Ralston}}{1999}]{Jain1999}
{Jain} P.,  {Ralston} J.~P.,  1999, \mn@doi [Modern Physics Letters A]
  {10.1142/S0217732399000481}, \href
  {http://adsabs.harvard.edu/abs/1999MPLA...14..417J} {14, 417}

\bibitem[\protect\citeauthoryear{{Jain}, {Narain}  \& {Sarala}}{{Jain}
  et~al.}{2004}]{Jain2004}
{Jain} P.,  {Narain} G.,   {Sarala} S.,  2004, \mn@doi [Mon. Not. R. Astron.
  Soc.] {10.1111/j.1365-2966.2004.07169.x}, \href
  {http://adsabs.harvard.edu/abs/2004MNRAS.347..394J} {347, 394}

\bibitem[\protect\citeauthoryear{{Kim} \& {Naselsky}}{{Kim} \&
  {Naselsky}}{2010}]{Kim2010}
{Kim} J.,  {Naselsky} P.,  2010, \mn@doi [Astrophysical Journal Letters]
  {10.1088/2041-8205/714/2/L265}, \href
  {http://adsabs.harvard.edu/abs/2010ApJ...714L.265K} {714, L265}

\bibitem[\protect\citeauthoryear{{Kim} \& {Naselsky}}{{Kim} \&
  {Naselsky}}{2011}]{Kim2011}
{Kim} J.,  {Naselsky} P.,  2011, \mn@doi [Astrophysical Journal]
  {10.1088/0004-637X/739/2/79}, \href
  {http://adsabs.harvard.edu/abs/2011ApJ...739...79K} {739, 79}

\bibitem[\protect\citeauthoryear{{Land} \& {Magueijo}}{{Land} \&
  {Magueijo}}{2005}]{Land2005}
{Land} K.,  {Magueijo} J.,  2005, \mn@doi [Phys. Rev. D]
  {10.1103/PhysRevD.72.101302}, \href
  {http://adsabs.harvard.edu/abs/2005PhRvD..72j1302L} {378, 153}

\bibitem[\protect\citeauthoryear{{Lewis}, {Challinor}  \& {Lasenby}}{{Lewis}
  et~al.}{2000}]{Lewis2000}
{Lewis} A.,  {Challinor} A.,   {Lasenby} A.,  2000, \mn@doi [Astrophysical Journal]
  {10.1086/309179}, \href {http://adsabs.harvard.edu/abs/2000ApJ...538..473L}
  {538, 473}
  
\bibitem[\protect\citeauthoryear{{Nadathur}, {Lavinto}, {Hotchkiss}, \& {R{\"a}s{\"a}nen}}{{Nadathur}
  et~al.}{2014}]{Nadathur2014}
{Nadathur} S., {Lavinto} M., {Hotchkiss} S., {R{\"a}s{\"a}nen} S., 2014,
  \mn@doi [Phys. Rev. D] {10.1103/PhysRevD.90.103510},
  \href {http://adsabs.harvard.edu/abs/2014PhRvD..90j3510N} {90, 103510}
  
\bibitem[\protect\citeauthoryear{{Naselsky}, {Christensen}, {Coles}, {Verkhodanov},
 {Novikov}, \& {Kim}}{{Naselsky} et~al.}{2010}]{Naselsky2010}
 {Naselsky} P.~D., {Christensen} P.~R., {Coles} P., {Verkhodanov} O.~V.,
 {Novikov} D.~I. {Kim} J., 2010,
  \mn@doi [Astrophys. Bull.]  {10.1134/S199034131002001X},
  \href {http://adsabs.harvard.edu/abs/2010AstBu..65..101N} {65, 101}

\bibitem[\protect\citeauthoryear{{Planck Collaboration XXIII}}{{Planck
  Collaboration XXIII}}{2013}]{Planck2013iso}
{Planck Collaboration XXIII}, 2013, \mn@doi [Astronomy and Astrophysics]
  {10.1051/0004-6361/201321534}, \href
  {http://adsabs.harvard.edu/abs/2014A%26A...571A..23P} {571, A23}

\bibitem[\protect\citeauthoryear{{Planck Collaboration IX}}{{Planck
  Collaboration IX}}{2016}]{Planck2015cmb}
{Planck Collaboration IX}, 2016, \mn@doi [Astronomy and Astrophysics]
  {10.1051/0004-6361/201525936}, \href
  {http://adsabs.harvard.edu/abs/2016A%26A...594A..42P} {594, A9}

\bibitem[\protect\citeauthoryear{{Planck Collaboration XI}}{{Planck
  Collaboration XI}}{2016}]{Planckpara2016a}
{Planck Collaboration XI}, 2016, \mn@doi [Astronomy and Astrophysics]
  {10.1051/0004-6361/201526926}, \href
  {http://adsabs.harvard.edu/abs/2016A%26A...594A..11P} {594, A11}

\bibitem[\protect\citeauthoryear{{Planck Collaboration XIII}}{{Planck
  Collaboration XIII}}{2016}]{Planckpara2016b}
{Planck Collaboration XIII}, 2016, \mn@doi [Astronomy and Astrophysics]
  {10.1051/0004-6361/201525830}, \href
  {http://adsabs.harvard.edu/abs/2016A%26A...594A..13P} {594, A13}

\bibitem[\protect\citeauthoryear{{Planck Collaboration XVI}}{{Planck
  Collaboration XVI}}{2016}]{Planck2015iso}
{Planck Collaboration XVI}, 2016, \mn@doi [Astronomy and Astrophysics]
  {10.1051/0004-6361/201526681}, \href
  {http://adsabs.harvard.edu/abs/2016A%26A...594A..16P} {594, A16}

\bibitem[\protect\citeauthoryear{{Prunet}, {Uzan}, {Bernardeau}  \&
  {Brunier}}{{Prunet} et~al.}{2005}]{Prunet2005}
{Prunet} S.,  {Uzan} J.~P.,  {Bernardeau} F.,   {Brunier} T.,  2005, \mn@doi
  [Phys. Rev. D] {10.1103/PhysRevD.71.083508}, \href
  {http://adsabs.harvard.edu/abs/2005PhRvD..71h3508P} {71, 083508}

\bibitem[\protect\citeauthoryear{{Ralston} \& {Jain}}{{Ralston} \&
  {Jain}}{2004}]{Ralston2004}
{Ralston} J.~P.,  {Jain} P.,  2004, \mn@doi [International Journal of Modern
  Physics D] {10.1142/S0218271804005948}, \href
  {http://adsabs.harvard.edu/abs/2004IJMPD..13.1857R} {13, 1857}

\bibitem[\protect\citeauthoryear{{Rath} \& {Jain}}{{Rath} \&
  {Jain}}{2013}]{Rath2013}
{Rath} P.~K.,  {Jain} P.,  2013, \mn@doi [Journal of Cosmology and
  Astroparticle Physics] {10.1088/1475-7516/2013/12/014}, \href
  {http://adsabs.harvard.edu/abs/2013JCAP...12..014R} {12, 14}

\bibitem[\protect\citeauthoryear{{Rath}, {Mudholkar}, {Jain}, {Aluri}  \&
  {Panda}}{{Rath} et~al.}{2013}]{Pranati2013}
{Rath} P.~K.,  {Mudholkar} T.,  {Jain} P.,  {Aluri} P.~K.,   {Panda} S.,  2013,
  \mn@doi [Journal of Cosmology and Astroparticle Physics]
  {10.1088/1475-7516/2013/04/007}, \href
  {http://adsabs.harvard.edu/abs/2013JCAP...04..007R} {4, 7}

\bibitem[\protect\citeauthoryear{{Rath}, {Aluri}  \& {Jain}}{{Rath}
  et~al.}{2015}]{Pranati2015}
{Rath} P.~K.,  {Aluri} P.~K.,   {Jain} P.,  2015, \mn@doi [Phys. Rev. D]
  {10.1103/PhysRevD.91.023515}, \href
  {http://adsabs.harvard.edu/abs/2015PhRvD..91b3515R} {91, 023515}

\bibitem[\protect\citeauthoryear{{Rath} \& {Samal}}{{Rath} \&
  {Samal}}{2015}]{Rath2015}
{Rath} P.~K.,  {Samal} P.~K.,  2015, \mn@doi [Modern Physics Letters A]
  {10.1142/S021773231550131X}, \href
  {http://adsabs.harvard.edu/abs/2015MPLA...3050131R} {30, 1550131}

\bibitem[\protect\citeauthoryear{{Samal}, {Saha}, {Jain}  \& {Ralston}}{{Samal}
  et~al.}{2008}]{Samal2008}
{Samal} P.~K.,  {Saha} R.,  {Jain} P.,   {Ralston} J.~P.,  2008,
  \mn@doi [Mon. Not. R. Astron. Soc.] {10.1111/j.1365-2966.2008.12960.x}, \href
  {http://adsabs.harvard.edu/abs/2008MNRAS.385.1718S} {385, 1718}

\bibitem[\protect\citeauthoryear{{Samal}, {Saha}, {Jain}  \& {Ralston}}{{Samal}
  et~al.}{2009}]{Samal2009}
{Samal} P.~K.,  {Saha} R.,  {Jain} P.,   {Ralston} J.~P.,  2009, \mn@doi [Mon.
  Not. R. Astron. Soc.] {10.1111/j.1365-2966.2009.14728.x}, \href
  {http://adsabs.harvard.edu/abs/2009MNRAS.396..511S} {396, 511}

\bibitem[\protect\citeauthoryear{{Samal}, {Saha}, {Delabrouille}, {Prunet},
  {Jain}  \& {Souradeep}}{{Samal} et~al.}{2010}]{Samal2010}
{Samal} P.~K.,  {Saha} R.,  {Delabrouille} J.,  {Prunet} S.,  {Jain} P.,
  {Souradeep} T.,  2010, \mn@doi [Astrophysical Journal]
  {10.1088/0004-637X/714/1/840}, \href
  {http://adsabs.harvard.edu/abs/2010ApJ...714..840S} {714, 840}

\bibitem[\protect\citeauthoryear{{Schwarz}, {Starkman}, {Huterer}  \&
  {Copi}}{{Schwarz} et~al.}{2004}]{Schwarz2004}
{Schwarz} D.~J.,  {Starkman} G.~D.,  {Huterer} D.,   {Copi} C.~J.,  2004,
  \mn@doi [Physical Review Letters] {10.1103/PhysRevLett.93.221301}, \href
  {http://adsabs.harvard.edu/abs/2004PhRvL..93v1301S} {93, 221301}

\bibitem[\protect\citeauthoryear{{Vielva}, {Mart{\'{\i}}nez-Gonz{\'a}lez},
 {Barreiro}, {Sanz}, \& {Cay{\'o}n}}{{Vielva} et~al.}{2004}]{Vielva2004}
{Vielva} P., {Mart{\'{\i}}nez-Gonz{\'a}lez} E., {Barreiro} R.~B.,
	{Sanz} J.~L., {Cay{\'o}n} L.,  2004,
  \mn@doi [Astrophysical Journal] {10.1086/421007}, \href
  {http://adsabs.harvard.edu/abs/2004ApJ...609...22V} {609, 22}

\bibitem[\protect\citeauthoryear{{Zaldarriaga} \& {Seljak}}{{Zaldarriaga} \&
  {Seljak}}{1997}]{Zaldarriaga1997}
{Zaldarriaga} M.,  {Seljak} U.,  1997, \mn@doi [Phys. Rev. D]
  {10.1103/PhysRevD.55.1830}, \href
  {http://adsabs.harvard.edu/abs/1997PhRvD..55.1830Z} {55, 1830}

\bibitem[\protect\citeauthoryear{{Zhao}}{{Zhao}}{2013}]{Zhao2013}
{Zhao} W.,  2013, \mn@doi [Mon. Not. R. Astron. Soc.] {10.1093/mnras/stt979}, \href
  {http://adsabs.harvard.edu/abs/2013MNRAS.433.3498Z} {433, 3498}

\bibitem[\protect\citeauthoryear{{Zhao}}{{Zhao}}{2014}]{Zhao2014}
{Zhao} W.,  2014, \mn@doi [Phys. Rev. D] {10.1103/PhysRevD.89.023010}, \href
  {http://adsabs.harvard.edu/abs/2014PhRvD..89b3010Z} {89, 023010}

\makeatother
\end{thebibliography}
\end{document}